\documentclass[a4paper,11pt,prd,preprint, amsfonts,amsmath,amssymb
]{revtex4}
\usepackage{bm}
\usepackage[dvipdfmx]{graphicx}
\usepackage{ascmac}
\usepackage{amsthm}
\usepackage{latexsym}
\usepackage{physics}
\usepackage{color,float}
\usepackage{hyperref}
\usepackage{mathtools}
\usepackage{simplewick}
\allowdisplaybreaks

\begin{document}
\title{Markovian quantum master equation with Poincar\'{e} symmetry}

\author{Kaito Kashiwagi} 
\email{kashiwagi.kaito.268@s.kyushu-u.ac.jp}
\author{Akira Matsumura}
\email{matsumura.akira@phys.kyushu-u.ac.jp}
\affiliation{Department of Physics, Kyushu University, Fukuoka, 819-0395, Japan}
\begin{abstract}
We investigate what kind of Markovian quantum master equation (QME) in the Gorini-Kossakowski-Sudarshan-Lindblad (GKSL) form is realized under Poincar\'{e} symmetry. 
The solution of the Markovian QME 
is given by a quantum dynamical semigroup, 
for which we introduce invariance under Poincar\'{e} transformations.
Using the invariance of the dynamical semigroup and applying the unitary representation of Poincar\'{e} group, we derive the Markovian QME for a relativistic massive spin-0 particle. 
Introducing the field operator of the massive particle and examining its evolution, we find that the field follows a dissipative Klein-Gordon equation. 
In addition, we show that any two local operators for spacelike separated regions commute with each other. 
This means that the microcausality condition is satisfied for the dissipative model of the massive particle.
\end{abstract}
\maketitle
\tableofcontents

\section{Introduction}

Markovian quantum master equation (QME) is of extensive importance in modern physics. 
A typical study using Markovian QME is on the dynamics of a quantum system coupled to its surrounding environment \cite{Breuer2002, Davies1976}.
Such a quantum system is called an open quantum system, and its Markovian process is governed by the Markovian QME. 
In the theory of open quantum systems, the celebrated Gorini-Kossakowski-Sudarshan-Lindblad (GKSL) form of the Markovian QME \cite{Gorini1976, Lindblad1976} is crucial for describing the decoherence and dissipative phenomena of an open quantum system. 
The description based on the GKSL form has been applied in quantum information science \cite{Breuer2002, Gardinar2000,  Nielsen2002, Kraus2008, Wiseman2010}.

Another study applying Markovian QME is on the interplay of quantum and gravity physics.
In modern physics, it has not been elucidated whether gravity follows quantum mechanics. 
This stimulates two theoretical standpoints: one is that ``gravity is quantum" and the other is that ``gravity is classical".
In the former standpoint, perturbative quantum gravity \cite{Donoghue1994} was established as an effective theory of quantum gravity. 
In this effective theory, we can describe the perturbative dynamics of canonically quantized matters and gravitational fields.
In the latter standpoint, the Kafri-Taylor-Milburn model \cite{Kafri2014} and the Di\'{o}si-Penrose or the Di\'{o}si-Tilloy model \cite{Diosi1987, Diosi1989, Penrose1996, Tilloy2016} were proposed. 
In these model,  gravity does not obey the principle of quantum superposition and is assumed to be in a classically definite state.  
Such gravity causes the collapse of the wave function of matter, which is described by a non-relativistic model based on a Markovian QME.

As mentioned so far, 
Markovian QME has been broadly used for the description of dissipative and collapse dynamics. 
In these contexts, it is intriguing to discuss the Markovian QME for the dissipative and collapse dynamics of relativistic quantum systems.
For example, the decay process of an unstable particle, such as pion decay, is considered to be a relativistic dissipative phenomenon. 
This phenomenon would be governed by a Markovian QME for a long time scale at which the Markovian approximation is valid. 
Such an unstable particle might be described by relativistic theories of open quantum fields \cite{Baidya2017, Meng2021, Wang2022}. Within the gravity-induced collapse of wave function, a relativisitic model using a Markovian (or possibly non-Markovian) QME should be preferred if the collapse occurs at a fundamental level and for a relativistic quantum matter. 
In Ref.\cite{Oppenheim2022}, for formulating such a model, the covariant theory of genuinely classical gravity coupled to quantum matters was discussed.
Also, regarding collapse dynamics not directly related to gravity, relativistic collapse models have been discussed in the literature \cite{Bedingham2014,Bedingham2019, Jones2021a, Jones2021b, Pearle2015, Kurkov2011}. 

Towards understanding the dissipative and collapse dynamics of relativistic quantum systems, in this paper, we discuss a relativistic Markovian QME in the GKSL form.
This equation is developed by adopting Poincar\'{e} symmetry as a guiding principle, which is commonly employed in relativistic theories, such as quantum field theories and perturbative quantum gravity in the Minkowski background.
To respect the symmetry, we define the Poincar\'{e} invariance of the quantum dynamical semigroup that gives the solution of Markovian QME.
In the previous work \cite{Matsumura2023}, the Poincar\'{e} invariant reduced dynamics of a massive spin-0 particle was discussed. 
However, its time-local evolution equation based on Markovian QME was not fully understood. 
In the present paper, leveraging the Poincar\'{e} invariance of the dynamical semigroup and the unitary representation of the Poincar\'{e} group, we construct the Markovian QME of a relativistic massive spin-0 particle. 
We further investigate the evolution of the particle field and assess whether the evolution is consistent with microcausality. 
We find that the field obeys a Klein-Gordon (KG) equation with a dissipative term and the microcausality is satisfied. 
This suggests that our approach has the potential to provide dissipative quantum field theories.

The structure of this paper is as follows. 
In Sec.II, we explain the basics of Markovian QME in the GKSL form. In Sec.III, we introduce the Poincar\'{e} group, its unitary representation, and the Poincar\'{e} algebra. 
We also define the condition that a Markovian QME has Poincaré symmetry, which is crucial for formulating the relativistic theory of Markovian QME. 
In Sec. IV we present the concrete model of a massive particle of spin 0 and investigate the field dynamics of the particle. 
We then find the evolution equation of the field and that the microcausality is satisfied. In Sec.V, we explain the derivation of the Markovian QME of the massive particle. In Sec.VI, the conclusion and the future prospects of this paper are devoted. 
Throughout this paper, we use the natural unit 
$\hbar=c=1$ and adopt the convention of the Minkowski metric as
$\eta_{\mu \nu}=\text{diag}[-1,1,1,1]$ used for lowering and raising indices. 
Also, the commutator and the anticommutator are defined as 
$[\hat{A},\hat{B}]=\hat{A}\hat{B}-\hat{B}\hat{A}$ and
$\{\hat{A},\hat{B}\}=\hat{A}\hat{B}+\hat{B}\hat{A}$, respectively. 

\section{Markovian quantum master equation}

Our main purpose is to formulate Markovian QMEs applicable to relativistic quantum systems. 
To this end, in this section, we introduce the Markovian QME in a general GKSL form \cite{Breuer2002,Davies1976, Gorini1976, Lindblad1976}, which can describe the dissipative dynamics of open quantum system and the collapse of the wave function of quantum system. 
The Markovian QME for the density operator 
$\rho(t)$ of a quantum system at time 
$t$ is
\begin{equation}
\frac{d}{dt}\rho(t)=\mathcal{L}[\rho(t)],
   \label{eq:rhot}
\end{equation}
where the GKSL generator 
$\mathcal{L}$ is defined as 
\begin{equation}
\mathcal{L}[\rho(t)]
= -i\left[ \hat{M}, \rho(t) \right] +  \sum _{\lambda} \left[ \hat{L}_{\lambda} \rho(t) \hat{L}^{\dagger}_{\lambda} - \frac{1}{2} \left\{ \hat{L}^{\dagger}_{\lambda} \hat{L}_{\lambda} , \rho(t) \right\} \right] 
\label{eq:L}
\end{equation}
with a Hermitian operator
$\hat{M}$ and noise operators 
$\hat{L}_{\lambda}$ called Lindblad operators. 
In Eq.\eqref{eq:L}, the first term describes the unitary evolution of the system, and the second term leads to the non-unitary evolution of the system, such as dissipation and wave function collapse.  
The GKSL generator 
$\mathcal{L}$ does not change under the following transformations \cite{Breuer2002,Holevo1998},
\begin{align}
&\hat{L}_{\lambda} \longrightarrow \hat{L}'_{\lambda} = \sum_{\lambda'} \mathcal{U}_{\lambda\lambda'} \hat{L}_{\lambda'} + \alpha_{\lambda}\hat{\mathbb{I}}, 
\label{eq:trans1}
\\
&\hat{M} \longrightarrow \hat{M}' = \hat{M} + \frac{1}{2i}\sum_{\lambda,\lambda'}\left[ \alpha^{\ast}_{\lambda}\mathcal{U}_{\lambda\lambda'}\hat{L}_{\lambda'} - \alpha_{\lambda}\mathcal{U}^{\ast}_{\lambda\lambda'}\hat{L}_{\lambda'}^{\dagger} \right] + \beta\hat{\mathbb{I}}, 
\label{eq:trans2}   
\end{align}
where $\alpha_{\lambda}$ are complex number, $\beta$ is real and $\mathcal{U}_{\lambda\lambda'}$ is unitary matrix satisfying 
$\sum_{\lambda} \mathcal{U}^*_{\lambda_{1}\lambda}\mathcal{U}_{\lambda_{2}\lambda} = \sum_{\lambda} \mathcal{U}^*_{\lambda\lambda_{1}}\mathcal{U}_{\lambda\lambda_{2}} = \delta_{\lambda_{1},\lambda{2}}$.
This invariance is important in discussing Poincar\'{e} symmetry in the later sections.
The formal solution of the QME \eqref{eq:rhot} is
\begin{equation}
\rho(t) = e^{\mathcal{L}t} \left[ \rho(0) \right],
\label{eq:eLt}
\end{equation}
where 
$e^{\mathcal{L}t}$ maps the initial state $\rho(0)$ onto the evolved state $\rho(t)$. 
The map
$e^{\mathcal{L}t}$ has two properties, complete positivity and trace-preserving, and the one parameter family $\{ e^{\mathcal{L}t}: t \geq0 \}$
is called a quantum dynamical semigroup\cite{Breuer2002}. 
In the next section, we will discuss Poincar\'{e} symmetry and introduce the Poincar\'{e} invariance for the quantum dynamical semigroup considered here. 
The invariance condition will be used to identify the Markovian QME of a massive spin-0 particle in Sec.IV. 

\section{Dynamical semigroup with Poincar\'{e} symmetry}

To make our formulation clear, in this section, we first explain the Poincar\'{e} symmetry in quantum theory \cite{Weinberg1995} and then define the Poincar\'{e} invariance of quantum dynamical semigroup. 
The Poincar\'{e} symmetry is respected for relativistic theories and requires invariance under the coordinate transformation 
$x^{\mu} \rightarrow x'^{\mu} = \Lambda^{\mu}_{\ \nu}x^{\nu} + a^{\mu}$, where 
$\Lambda^{\mu}{}_{\nu}$ give the Lorentz transformation matrix satisfying  
$\eta_{\mu\nu}=\eta_{\rho \sigma} \Lambda^{\rho}{}_\mu \Lambda^{\sigma}{}_\nu$ and 
$a^{\mu}$ are the parameters of spacetime translation.
This transformation is called a Poincar\'{e} transformation and  preserves the line element of the Minkowski spacetime, 
$ds^2=\eta_{\mu\nu} dx^\mu dx^\nu$. 
In the following, we only consider the continuous transformation, that is, the proper orthochronous Lorentz transformation with 
$\Lambda^0{}_0 \geq 1$ and 
$\det \Lambda=1$. 

Denoting the Poincar\'{e} transformation as the map
$g(\Lambda, a): x^{\mu} \rightarrow x'^{\mu} = \Lambda^{\mu}_{\ \nu}x^{\nu} + a^{\mu}$, this map satisfies the multiplication rule
\begin{equation}
g(\Lambda', a')g(\Lambda, a) = g(\Lambda'\Lambda, \Lambda'a + a'). 
\label{eq:gg}
\end{equation}
From Eq.\eqref{eq:gg}, the inverse transformation and the identical transformation can be given as 
$g(I,0)$ and $g(\Lambda^{-1}, - \Lambda^{-1}a)$, respectively, where  
$I$ is the identity matrix and 
$\Lambda^{-1}$ is the inverse of the Lorentz transformation matrix 
$\Lambda$. 
Therefore, the whole of Poincar\'{e} transformation 
$g(\Lambda, a)$ forms a group, which is called Poincar\'{e} group.

In quantum theory, the Poincar\'{e} transformation 
$g(\Lambda, a)$ is represented on a Hilbert space as a unitary operator 
$\hat{U}(\Lambda, a)$ satisfying 
\begin{equation}
\hat{U}(\Lambda', a')\hat{U}(\Lambda, a)=\hat{U}(\Lambda'\Lambda, \Lambda'a + a').
\label{eq:UU}
\end{equation}
This unitary operator 
$\hat{U}(\Lambda, a)$ is called the unitary representation of Poincar\'{e} group.
Considering the infinitesimal Poincar\'{e} transformation 
$g(I+\omega,\epsilon)$  with
\begin{equation}
\Lambda^{\mu}_{\ \nu} = \delta^{\mu}_{\ \nu} + \omega^{\mu}_{\ \nu}, \quad a^{\mu} = \epsilon^{\mu}, 
\label{eq:inf.trans.}
\end{equation}
where 
$\omega^{\mu}_{\ \nu}$ with 
$\omega_{\mu \nu}=-\omega_{\nu \mu}$ and
$\epsilon^{\mu}$ are infinitesimal parameters, and $\delta^{\mu}_{\ \nu}$ is Kronecker's symbol, we can expand the unitary operator 
$\hat{U}(I + \omega, \epsilon)$ as
\begin{equation}
\hat{U}(I + \omega, \epsilon) 
= \hat{I} + \frac{i}{2}\omega_{\mu\nu}\hat{J}^{\mu\nu} -i \epsilon_{\mu}\hat{P}^{\mu} + (\mathrm{high \ order \ term \ with \ respect \ to \ } \omega \ \mathrm{and} \ \epsilon), \label{eq:inf.U}
\end{equation}
where the Hermitian operators
$\hat{J}^{\mu\nu}$ with 
$\hat{J}^{\mu\nu}=-\hat{J}^{\nu\mu}$ and  
$\hat{P}^{\mu}$ in the Schr\"{o}dinger picture are the generators of Lorentz transformation and spacetime translation, respectively\cite{Weinberg1995}. 
These generators satisfy the following commutation relations,
\begin{align}
\left[ \hat{J}_k, \hat{J}_\ell \right] &= i\epsilon_{k\ell m}\hat{J}^{m}, 
\label{eq:JJ}
\\
\left[ \hat{J}_k, \hat{K}_\ell \right] &= i\epsilon_{k\ell m}\hat{K}^{m},
\label{eq:JK}
\\
\left[ \hat{K}_i, \hat{K}_j \right] &= - i \epsilon_{ijk}\hat{J}^{k},
\label{eq:KK}, 
\\
\left[\hat{J}_k, \hat{P}_\ell \right] &= i\epsilon_{k\ell m}\hat{P}^{m}, 
\label{eq:JP}
\\
\left[ \hat{K}_j, \hat{H} \right] &=i\hat{P}_j
\label{eq:KH}
\\
\left[ \hat{K}_i, \hat{P}_j \right] &= i \delta_{ij}\hat{H}
\label{eq:KP}
\\
\left[ \hat{P}_i, \hat{P}_j \right] &= 0
\label{eq:PP}
\\
\left[ \hat{P}_i , \hat{H} \right]&=0=\left[ \hat{J}_{i} , \hat{H} \right] 
\label{eq:PHJH}
\end{align}
where we used the notations 
$\hat{P}^{\mu} = ( \hat{H}, \hat{P}^{1}, \hat{P}^{2}, \hat{P}^{3} )$, 
$\hat{J}^{k} = \frac{1}{2} \epsilon^{ijk} \hat{J}_{ij} = (\hat{J}^{23}, \hat{J}^{31}, \hat{J}^{12})$, and $\hat{K}^{k} = \hat{J}^{k0} = ( \hat{J}^{10}, \hat{J}^{20}, \hat{J}^{30} )$. 
The above commutation relations are called the Poincar\'{e} algebra. 
From this Poincar\'{e} algebra, it turns out that the boost operators 
$\hat{K}^{k}$ do not commute with the Hamiltonian 
$\hat{H}$. 
In relativistic unitary theories such as quantum field theories, the boost operator in the Heisenberg picture, $\hat{K}^{k}_\text{H} = e^{i\hat{H}t} \hat{K}^{i} e^{-i\hat{H}t}$, is conserved by the Noether theorem. 
This leads to that 
$\hat{K}^{k}$ in the Schr\"{o}dinger picture is  $\hat{K}^{i}_{0} - t\hat{P}^{i}$, where 
$\hat{K}^i_0$ is 
$\hat{K}^i$ at 
$t=0$. 
Hence the unitary
$\hat{U}(\Lambda,a)$ has the explicit time dependence in the Schr\"{o}dinger picture. 
In the following, we denote 
$\hat{K}^i$ and 
$\hat{U}(\Lambda,a)$ as 
$\hat{K}^i_t$ and 
$\hat{U}_t (\Lambda,a)$, respectively, to specify the time dependence in the Schr\"{o}dinger picture.

We are in a position to define the Poincar\'{e} invariance of quantum dynamical semigroup. 
If the dynamical semigroup 
$\{e^{\mathcal{L}t} : t\geq 0\} $ given in \eqref{eq:eLt} satisfies the following condition
\begin{equation}
\hat{U}_{t}(\Lambda, a) e^{\mathcal{L}t} \left[ \rho(0) \right] \hat{U}^{\dagger}_{t}(\Lambda, a) = e^{\mathcal{L}t} \left[ \hat{U}_{0}(\Lambda, a) \rho(0) \hat{U}^{\dagger}_{0}(\Lambda, a) \right], 
\label{eq:Pinv}
\end{equation}
we regard that 
$\{e^{\mathcal{L}t} : t\geq 0\} $ has the Poincar\'{e} invariance (see also \cite{Matsumura2023}).
The similar invariance condition for symmetry group has been discussed in Refs.\cite{Cirstoiu2020, Marvian2014, Marko2017}. 
The unitary operator 
$\hat{U}_{t}(\Lambda, a)$ is generated by $\hat{P}^{\mu}$, $\hat{J}^{k}$, and $\hat{K}^{i}_t = \hat{K}^{i}_{0} - t\hat{P}^{i}$, which satisfies the group multiplication rule 
$\hat{U}_{t}(\Lambda', a')\hat{U}_{t}(\Lambda, a)=\hat{U}_{t}(\Lambda'\Lambda, \Lambda'a + a')$.
In unitary relativistic theories, the time evolution of a density operator  
$\rho(t)$ is given as
\begin{equation}
\rho(t) = \mathcal{U}_{t} [ \rho(0) ] = e^{-i\hat{H}t} \rho(0) e^{i\hat{H}t}, 
\label{eq:calUt}
\end{equation}
and this map 
$\mathcal{U}_{t}$ satisfies the condition Eq.\eqref{eq:Pinv} as
\begin{equation}
\hat{U}_{t}(\Lambda, a) \mathcal{U}_{t} \left[ \rho(0) \right] \hat{U}^{\dagger}_{t}(\Lambda, a) = \mathcal{U}_{t} \left[ \hat{U}_{0}(\Lambda, a) \rho(0) \hat{U}^{\dagger}_{0}(\Lambda, a) \right].
\label{eq:Uinv}
\end{equation}
So, the invariance condition \eqref{eq:Pinv} for 
$\{e^{\mathcal{L}t} : t\geq 0\}$ is a generalization of that for $\{ \mathcal{U}_t : t\geq 0\}$. 
We can get the derivative form of the invariance condition \eqref{eq:Pinv}. 
Differentiating the both sides of the condition \eqref{eq:Pinv} with respect to time $t$, we obtain
\begin{equation}
\frac{d}{dt} \rho'(t) = \mathcal{L}\rho'(t), 
\label{eq:drho'/dt}
\end{equation}
where 
$\rho'(t) = \hat{U}_t (\Lambda,a)\rho(t)\hat{U}^{\dagger}_t (\Lambda,a)$. 
Eq.\eqref{eq:drho'/dt} means that the generator of the time evolution, that is, the GKSL generator 
$\mathcal{L}$ is invariant under the Poincar\'{e} transformation  
$\rho(t) \rightarrow \rho'(t)$. 

Here, we explain our motivation why the Poincar\'{e} invariance is adopted.
In the context of dissipative dynamics, the decay of an unstable particle is a relativistic dissipative phenomenon. 
In the context of collapse dynamics, possibly induced by gravity, if the collapse of wave function is fundamental, its relativistic description should be prefered.
These dissipative and collapse dynamics may be relativistically symmetric. 
As a step to exploring this possibility, in this paper, we are considering a framework with the Poincar\'{e} symmetry as a relativistic symmetry. 

To get a simple invariance condition, we reduce Eq. \eqref{eq:Pinv} for 
$\{e^{\mathcal{L}t} : t\geq 0\} $ to the condition for the GKSL generator
$\mathcal{L}$. 
Evaluating Eq.\eqref{eq:Pinv} for a small $t$ with the help of 
\begin{equation}
\hat{K}^i_t = e^{-i\hat{H}t} \hat{K}^{i}_{0} e^{i\hat{H}t} 
\label{eq:Kt}
\end{equation}
and 
\begin{equation}
\hat{U}_{t}(\Lambda,a) = e^{-i\hat{H}t} \hat{U}_{0}(\Lambda,a) e^{i\hat{H}t},
\label{eq:Ut}
\end{equation}
we obtain 
\begin{equation}
\hat{U}_{0}(\Lambda,a)\left( \mathcal{L} - \mathcal{H} \right)\left[ \rho(0) \right] \hat{U}^{\dagger}_{0}(\Lambda,a) = \left( \mathcal{L} - \mathcal{H} \right) \left[ \hat{U}_{0}(\Lambda,a)\rho(0)\hat{U}^{\dagger}_{0}(\Lambda,a) \right], 
\label{eq:L-H}
\end{equation}
where 
$\mathcal{H}[\rho(t)]$ means 
$\mathcal{H}[\rho(t)] = -i [ \hat{H}, \rho(t)]$. 
Our task is to get a Markovian QME for relativistic quantum systems from the GKSL generator 
$\mathcal{L}$ which satisfies Eq.\eqref{eq:L-H}. 
In the next section, as an example of the Markovian QME derived from \eqref{eq:L-H}, we will present the Markovian QME of a massive relativistic particle of spin 0. 
After the next section, we will derive the QME.

\section{Model of a massive spin-0 particle and its field dynamics}
\label{sec:model}

In this section, we exemplify the Markovian QME of a massive spin-0 particle with a mass, whose the GKSL generator 
$\mathcal{L}$ satisfies\eqref{eq:L-H}. 
The Markovian QME is 
\begin{equation}
\frac{d}{dt}\rho(t) = \mathcal{H} [\rho(t)]+ \mathcal{D}[\rho(t)],
\label{eq:model}
\end{equation}
where the first term 
$\mathcal{H} [\rho(t)]= -i\left[\hat{H}, \rho(t)\right]$ is given by the free Hamiltonian,
\begin{equation}
\hat{H} = \int d^{3}p \ E_{\bm{p}}  \hat{a}^{\dagger}(\bm{p})\hat{a}(\bm{p}), 
\label{eq:H}
\end{equation}
with the energy of the particle, $E_{\bm{p}}=\sqrt{\bm{p}^{2}+m^{2}}$, and the second dissipation term is
\begin{equation}
\mathcal{D}[\rho(t)] = \gamma \int d^{3}p \left[ \hat{a}(\bm{p}) \rho(t) \hat{a}^{\dagger}(\bm{p})  - \frac{1}{2}\left\{\hat{a}^{\dagger}(\bm{p})\hat{a}(\bm{p}), \rho(t)\right\}\right]
\label{eq:D}
\end{equation}
with a non-negative dissipation rate 
$\gamma$. 
The superoperator 
$\mathcal{D}$ in \eqref{eq:D} is called a dissipator. 
The annihilation and creation operators of the massive particle,
$\hat{a}(\bm{p})$ and 
$\hat{a}^{\dagger}(\bm{p})$, satisfy the following commutation relations
\begin{equation}
\left[\hat{a}(\bm{p}), \hat{a}^{\dagger}(\bm{p}')\right] = \delta^{3}(\bm{p} - \bm{p}'), \quad  \left[ \hat{a}(\bm{p}), \hat{a}(\bm{p}') \right] = 0. 
\label{eq:aadg}
\end{equation}
It is easy to check the GKSL generator 
$\mathcal{L}=\mathcal{H}+\mathcal{D}$ 
of the Markovian QME \eqref{eq:model} satisfies the invariance condition \eqref{eq:L-H}, whose detail is provided in Appendix \ref{app:model}. 
The equation \eqref{eq:model} will be explicitly derived in the next section. 
 
In this section, we investigate the properties of the model by introducing the field operator of the massive particle,
\begin{equation}
\hat{\Phi}(\bm{x}) = \frac{1}{(2\pi)^{\frac{3}{2}}} \int d^{3}p \ \frac{1}{\sqrt{2E_{\bm{p}}}} \left( e^{i\bm{p}\cdot\bm{x}} \hat{a}(\bm{p}) + e^{-i\bm{p}\cdot\bm{x}} \hat{a}^{\dagger}(\bm{p}) \right), \label{eq:Phi}
\end{equation}
where 
$\bm{x}$ is a spatial coordinate. 
Let us first seek the time evolution equation of the massive spin-0 field in the Heisenberg picture. 
The time evolved field 
$\hat{\Phi}_\text{H}(t,\bm{x})$ is formally given as
\begin{equation}
\hat{\Phi}_\text{H}(t,\bm{x}) = e^{\mathcal{L}^{\dagger}t} \left[ \hat{\Phi}(\bm{x}) \right] 
, 
\label{eq:PhiH}
\end{equation}
where the adjoint of the GKSL generator 
$\mathcal{L}=\mathcal{H}+\mathcal{D}$, denoted by 
$\mathcal{L}^\dagger$, is defined as
\begin{equation}
\mathcal{L}^{\dagger} \left[ \hat{\Phi}_\text{H} \right]= \mathcal{H}^\dagger \left[ \hat{\Phi}_\text{H} \right] + \mathcal{D}^{\dagger}\left[ \hat{\Phi}_\text{H} \right], 
\label{eq:Ldg}
\end{equation}
with 
\begin{equation}
\mathcal{H}^\dagger[\hat{\Phi}_\text{H}]=i\left[\hat{H}, \hat{\Phi}_\text{H} \right], \quad 
\mathcal{D}^{\dagger}\left[ \hat{\Phi}_\text{H} \right] = \gamma \int d^{3}p  \left( \hat{a}^{\dagger}(\bm{p}) \hat{\Phi}_\text{H} \hat{a}(\bm{p}) - \frac{1}{2} \left\{ \hat{a}^{\dagger}(\bm{p})\hat{a}(\bm{p}), \hat{\Phi}_\text{H} \right\} \right). 
\label{eq:HDdg}
\end{equation}
For convenience, we consider the following operator $\hat{\Phi}_\text{I}(t,\bm{x})$ given as
\begin{equation}
\hat{\Phi}_\text{I}(t,\bm{x}) = e^{i\hat{H}t} \hat{\Phi}(\bm{x}) e^{-i\hat{H}t}. 
\label{eq:PhiI}    
\end{equation}
The relation between 
$\hat{\Phi}_\text{H}(t,\bm{x})$ and 
$\hat{\Phi}_\text{I}(t,\bm{x})$ is
\begin{equation}
\hat{\Phi}_\text{H}(t,\bm{x}) = e^{\mathcal{D}^{\dagger}t} \left[ \hat{\Phi}_\text{I}(t,\bm{x}) \right] 
.
\label{eq:HI}
\end{equation}
Calculating the operation
$\mathcal{D}^\dagger$ for 
$\hat{\Phi}_\text{I}(t,\bm{x})$, 
we get the time evolution 
$\hat{\Phi}_\text{H}(t,\bm{x})$ from Eq.\eqref{eq:HI}. 
Substituting \eqref{eq:Phi} into \eqref{eq:PhiI}, we obtain the operator 
$\hat{\Phi}_\text{I}(t,\bm{x})$ as  
\begin{equation}
\hat{\Phi}_\text{I}(t,\bm{x}) = \frac{1}{(2\pi)^{\frac{3}{2}}} \int  \frac{d^{3}p}{\sqrt{2E_{\bm{p}}}} \left( e^{i(\bm{p}\cdot\bm{x} - E_{\bm{p}}t)} \hat{a}(\bm{p}) + e^{-i(\bm{p}\cdot\bm{x} - E_{\bm{p}}t)} \hat{a}^{\dagger}(\bm{p}) \right), 
\label{eq:PhiI2}
\end{equation}
where we used the Baker-Campbell-Hausdorff formula and the commutation relations
$\left[ \hat{H}, \hat{a}^{\dagger}(\bm{p}) \right] = E_{\bm{p}}\hat{a}^{\dagger}(\bm{p})$ and 
$\left[ \hat{H}, \hat{a}(\bm{p}) \right] = - E_{\bm{p}}\hat{a}(\bm{p})$. 
We compute 
$\mathcal{D}^{\dagger}[\hat{\Phi}_\text{I}(t,\bm{x})]$ as 
\begin{align}
\mathcal{D}^{\dagger} \left[ \hat{\Phi}_\text{I}(t,\bm{x}) \right]
&= \gamma \int d^{3}p'  \left( \hat{a}^{\dagger}(\bm{p}') \hat{\Phi}_\text{I}(t,\bm{x}) \hat{a}(\bm{p}') - \frac{1}{2} \left\{ \hat{a}^{\dagger}(\bm{p}')\hat{a}(\bm{p}'), \hat{\Phi}_\text{I}(t,\bm{x}) \right\} \right)
\nonumber 
\\
&= \frac{\gamma}{2} \frac{1}{(2\pi)^{\frac{3}{2}}} \int \frac{d^{3}p}{\sqrt{2E_{\bm{p}}}} \left( e^{i(\bm{p}\cdot\bm{x} - E_{\bm{p}}t)} \left[\hat{N}, \hat{a}(\bm{p})\right]  + e^{-i(\bm{p}\cdot\bm{x} - E_{\bm{p}}t)} \left[ \hat{a}^{\dagger}(\bm{p}), \hat{N} \right] \right)
\nonumber 
\\
&= - \frac{\gamma}{2} \hat{\Phi}_\text{I}(t,\bm{x}),
\label{eq:comp.Ddg}
\end{align}
where 
$\hat{N}$ is the number operator defined as
\begin{equation}
\hat{N} = \int d^{3}p \hat{a}^{\dagger}(\bm{p})\hat{a}(\bm{p}). 
\label{eq:N}
\end{equation}
Repeating the operation of $\mathcal{D}^\dagger$ for $\hat{\Phi}_\text{I}(t,\bm{x})$, we get 
$\left( \mathcal{D}^{\dagger} \right)^{n} \left[ \hat{\Phi}_\text{I}(t,\bm{x}) \right] = \left( - \frac{\gamma}{2} \right)^{n} \hat{\Phi}_\text{I}(t,\bm{x})$ and hence 
\begin{equation}
\hat{\Phi}_\text{H}(t,\bm{x})= e^{\mathcal{D}^{\dagger}t} \left[ \hat{\Phi}_\text{I}(t,\bm{x}) \right] = \sum_{n=0}^{\infty} \frac{\left(\mathcal{D}^{\dagger}t\right)^{n}}{n!}\left[ \hat{\Phi}_\text{I}(t,\bm{x}) \right] = \sum_{n=0}^{\infty} \frac{1}{n!} \left( - \frac{\gamma}{2}t \right)^{n} \hat{\Phi}_\text{I}(t,\bm{x}) = e^{-\frac{\gamma}{2}t}\hat{\Phi}_\text{I}(t,\bm{x}). \label{eq:PhiH2}
\end{equation}
Noticing that the field operator 
$\hat{\Phi}_\text{I}(t,\bm{x})$ follows the usual KG equation, 
$\partial^2_t \hat{\Phi}_\text{I}(t,\bm{x})=(\nabla^2-m^2)\hat{\Phi}_\text{I}(t,\bm{x})$, we find that $\hat{\Phi}_\text{H}(t,\bm{x})$ obeys a dissipative KG equation,
\begin{equation}
\left[ \frac{\partial^{2}}{\partial t^{2}} + \gamma \frac{\partial}{\partial t} + \frac{\gamma^{2}}{4} \right] \hat{\Phi}_\text{H}(t,\bm{x}) = \left( \nabla^{2} - m^{2} \right) \hat{\Phi}_\text{H}(t,\bm{x}).
\label{eq:diss.KG}
\end{equation}
If the dissiapation rate vanishes,
$\gamma = 0$, \eqref{eq:diss.KG} becomes the usual KG equation. 
In the case where 
$\gamma \neq 0$, the term
$\gamma  \partial_t \hat{\Phi}_\text{H}$ represents the friction effect, which represents the dissipative dynamics of the field.

We examine whether microcausality holds in the dissipative dynamics.
The microcausality condition suggests that any two local operators at spacelike distance commute with each other. 
In special relativity, there is no transmission beyond the speed of light, and this fact is reflected in the notion of microcausality. 
Using the solution of the field 
$\hat{\Phi}_\text{H}(t,\bm{x})$, we can check that the field satisfies the microcausality as  
\begin{equation}
\left[ \hat{\Phi}_\text{H}(x^{0}, \bm{x}), \hat{\Phi}_\text{H}(y^{0}, \bm{y}) \right] = 0, 
\label{eq:PhiPhi}
\end{equation}
for any spacelike distance
$(x^{0} - y^{0})^{2} - (\bm{x} - \bm{y})^{2} < 0$. 
Here, we introduce a local operator 
\begin{equation}
\hat{\Pi}(\bm{x}) = - \frac{i}{(2\pi)^{\frac{3}{2}}} \int d^{3}p \sqrt{\frac{E_{\bm{p}}}{2}} \left[ e^{i \bm{p}\cdot\bm{x} } \hat{a}(\bm{p}) - e^{ - i \bm{p}\cdot\bm{x} } \hat{a}^{\dagger}(\bm{p}) \right],
\label{eq:Pi}
\end{equation}
and the evolved operator $\hat{\Pi}_\text{H}(t, \bm{x})=e^{\mathcal{L}^{\dagger}t}[\hat{\Pi}(\bm{x})]$ is given as
\begin{equation}
\hat{\Pi}_\text{H}(t, \bm{x}) = e^{ -\frac{\gamma}{2}t } \hat{\Pi}_\text{I}(t,\bm{x})
\label{eq:PiH}
\end{equation}
with 
\begin{align}
\hat{\Pi}_\text{I}(t, \bm{x})
& = \frac{\partial}{\partial t} \hat{\Phi}_\text{I}(t,\bm{x})
\nonumber 
\\
&= - \frac{i}{(2\pi)^{\frac{3}{2}}} \int d^{3}p \sqrt{\frac{E_{\bm{p}}}{2}} \left[ e^{i \left(\bm{p}\cdot\bm{x} -E_{\bm{p}}t \right) } \hat{a}(\bm{p}) - e^{ - i \left(\bm{p}\cdot\bm{x} -E_{\bm{p}}t \right) } \hat{a}^{\dagger}(\bm{p}) \right].
\label{eq:PiI}
\end{align}
For the field 
$\hat{\Phi}_\text{H}$ and the operator 
$\hat{\Pi}_\text{H}$, the simultaneous time commutation relation 
$[\hat{\Phi}_\text{H}(t, \bm{x}), \hat{\Pi}_\text{H}(t, \bm{y})]=ie^{-\gamma t} \delta^3(\bm{x}-\bm{y})$ holds. 
In this paper, the local operator 
$\hat{\Pi}_\text{H}$ is just called the conjugate momentum of $\hat{\Phi}_\text{H}$ even though these operators do not follow the usual canonical commutation relations.
We then find that the field 
$\hat{\Phi}_\text{H}$ and its conjugate momentum
$\hat{\Pi}_\text{H}$ satisfy
\begin{equation}
\left[ \hat{\Pi}_\text{H}(x^{0},\bm{x}), \hat{\Pi}_\text{H}(y^{0},\bm{y}) \right]=0=\left[\hat{\Phi}_\text{H}(x^{0}, \bm{x}), \hat{\Pi}_\text{H}(y^{0}, \bm{y})\right]
\label{eq:PiPhi}
\end{equation}
for $(x^{0} - y^{0})^{2} - (\bm{x} - \bm{y})^{2} < 0$. 
The detailed proof of the above three commutation relations are provided in Appendix \ref{sec:pf.comm.}. 
Furthermore, we can show that the results \eqref{eq:PhiPhi} and \eqref{eq:PiPhi} holds for more general local operators. 
As shown in Ref.\cite{Weinberg1995}, arbitrary operator of particle is expanded by the creation and annihilation operators of the particle.
This leads to that arbitrary local operator 
$\hat{O}$ at time 
$t=0$ for a finite spatial region O is expanded by 
$\hat{\Phi}(\bm{x})$ and 
$\hat{\Pi}(\bm{x})$, and the time evolution of 
$\hat{O}$ is given as 
\begin{align}
\hat{O}_\text{H}(t) 
&= \sum_{N, M} \int d^3x'_1 \cdots d^3x'_N \cdots d^3x_1 \cdots d^3x_M 
\sum_{i'_1, \cdots, i'_N = 1,2} \sum_{i_1, \cdots, i_M = 1,2} 
O^{i'_1 \cdots i'_N i_1 \cdots i_M}_{NM} (\bm{x}'_1, \cdots , \bm{x}'_N, \bm{x}_1, \cdots , \bm{x}_M) 
\nonumber
\\
&
\times 
e^{\mathcal{L}^{\dagger}t}
[\hat{\xi}^{i'_1}(\bm{x}'_{1}) \cdots \hat{\xi}^{i'_N}(\bm{x}'_N) \hat{\xi}^{i_1}(\bm{x}_1) \cdots \hat{\xi}^{i_M}(\bm{x}_M)], 
\label{eq:O} 
\end{align}
where the expansion coefficients
$O^{i'_1 \cdots i'_N i_1 \cdots i_M}_{NM} (\bm{x}'_1, \cdots , \bm{x}'_N, \bm{x}_1, \cdots , \bm{x}_M)$ only have nonzero values for the region O, and 
$\hat{\xi}^j (\bm{x})=\left(\hat{\Phi}(\bm{x}), \hat{\Pi}(\bm{x})\right)$.
Note that in the non-unitary theory considered here, $e^{\mathcal{L}t}[\hat{\xi}^{i'_1}(\bm{x}'_1) \cdots \hat{\xi}^{i'_N}(\bm{x}'_N) \hat{\xi}^{i_1}(\bm{x}_1) \cdots \hat{\xi}^{i_M}(\bm{x}_M)]
\neq \hat{\xi}_\text{H}^{i'_1}(t,\bm{x}'_1) \cdots \hat{\xi}_\text{H}^{i'_N}(t,\bm{x}'_N) \hat{\xi}_\text{H}^{i_1}(t,\bm{x}_1) \cdots \hat{\xi}_\text{H}^{i_M}(t,\bm{x}_M)$. 
In Appendix \ref{sec:prod}, we show that the time evolution of the product $\hat{\xi}^{i'_1}(\bm{x}'_1) \cdots \hat{\xi}^{i'_N}(\bm{x}'_N) \hat{\xi}^{i_1}(\bm{x}_1) \cdots \hat{\xi}^{i_M}(\bm{x}_M)$ is given as the product of operators chosen from $\hat{\xi}_\text{H}^{i'_1}(t,\bm{x}'_1), \dots, \hat{\xi}^{i'_N}_\text{H}(t,\bm{x}'_N), \hat{\xi}_\text{H}^{i_1}(t,\bm{x}_1), \dots, \hat{\xi}_\text{H}^{i_M}(t,\bm{x}_M)$.
This implies that the commutation relation of arbitrary local operators is computed from the commutation relations 
$[\hat{\xi}^i_\text{H}(x^0,\bm{x}), \hat{\xi}^j_\text{H}(y^0, \bm{y})]$, which vanish for 
$(x^0-y^0)^2-(\bm{x}-\bm{y})^2<0$ as shown in \eqref{eq:PhiPhi} and $\eqref{eq:PiPhi}$. 
Hence, the time-evolved operators 
$\hat{A}_\text{H} (t)$ and 
$\hat{B}_\text{H} (t')$ of
any two local operators $\hat{A}$ at time 
$t=0$ for a spatial region A  and 
$\hat{B}$ at time 
$t'=0$ for a spatial region B commutes if 
$\hat{A}_\text{H} (t)$ and 
$\hat{B}_\text{H} (t')$ are spacelike separated.
Therefore, the microcausality holds for the present model of the massive particle.  

In the following section, we will derive the Markovian QME of the massive particle considered here.

\section{Derivation of the Markovian quantum master equation}

In this section, we derive the Markovian QME of the massive spin-0 particle, which was presented in the previous section. 
For the derivation, let us first return to the condition Eq.\eqref{eq:L-H} of the Poincar\'{e} invariance, 
$\hat{U}_{0}(\Lambda,a)\left( \mathcal{L} - \mathcal{H} \right)\left[ \rho(0) \right] \hat{U}^{\dagger}_{0}(\Lambda,a) = \left( \mathcal{L} - \mathcal{H} \right) \left[ \hat{U}_{0}(\Lambda,a)\rho(0)\hat{U}^{\dagger}_{0}(\Lambda,a) \right]$.
Because the superoperator $\mathcal{L} - \mathcal{H}$ is given as
\begin{equation}
(\mathcal{L} - \mathcal{H})\left[ \rho \right] 
= -i \left[ \hat{M} - \hat{H}, \rho \right] + \sum_{\lambda} \left[ \hat{L}_{\lambda}\rho\hat{L}^{\dagger}_{\lambda} - \frac{1}{2} \left\{ \hat{L}^{\dagger}_{\lambda}\hat{L}_{\lambda}, \rho \right\} \right],
\label{eq:L-H2}
\end{equation}
the condition Eq.\eqref{eq:L-H} is rewritten as
\begin{align}
(\mathcal{L-H})[\rho] 
&= \hat{U}^{\dagger}_0 (\Lambda,a) \left( \mathcal{L} - \mathcal{H} \right) \left[ \hat{U}_{0}(\Lambda,a)\rho(0)\hat{U}^{\dagger}_0(\Lambda,a) \right] \hat{U}_{0}(\Lambda,a)
\nonumber
\\
&= -i \left[ \hat{U}^{\dagger}_0(\Lambda,a)( \hat{M} - \hat{H} )\hat{U}_0 (\Lambda,a), \rho(0) \right]
+
\sum_{\lambda} 
\Big[ 
\left( \hat{U}^{\dagger}_0 (\Lambda,a)\hat{L}_{\lambda}\hat{U}_{0}(\Lambda,a) \right) 
\rho(0) 
\left( \hat{U}^{\dagger}_0 (\Lambda,a) \hat{L}^{\dagger}_{\lambda}\hat{U}_0 (\Lambda,a)  \right) 
\nonumber 
\\
&-\frac{1}{2}\left\{  \left( \hat{U}^{\dagger}_{0}(\Lambda,a)\hat{L}_{\lambda}\hat{U}_{0}(\Lambda,a) \right)\left( \hat{U}^{\dagger}_{0}(\Lambda,a)\hat{L}^{\dagger}_{\lambda}\hat{U}_{0}(\Lambda,a)  \right), \rho(0) \right\} \Big], 
\label{eq:L-H3}
\end{align}
Eq.\eqref{eq:L-H3} means that the Markovian QME has invariance for the transformations $\hat{M} - \hat{H} \longrightarrow \hat{U}^{\dagger}_{0}(\Lambda,a)( \hat{M} - \hat{H} )\hat{U}_{0}(\Lambda,a)$ and $\hat{L}_{\lambda} \longrightarrow \hat{U}^{\dagger}_{0}(\Lambda,a)\hat{L}_{\lambda}\hat{U}_{0}(\Lambda,a)$.
Recalling that the form of Markovian QME does not change under the transformations 
\eqref{eq:trans1} and \eqref{eq:trans2}, we get the following rules, 
\begin{align}
&\hat{U}^{\dagger}_{0}(\Lambda,a)\hat{L}_{\lambda}(\Lambda,a)\hat{U}_{0}(\Lambda,a) = \sum_{\lambda'}\mathcal{U}_{\lambda\lambda'}(\Lambda,a)\hat{L}_{\lambda'} + \alpha_{\lambda}(\Lambda,a)\hat{\mathbb{I}},
\label{eq:UdgLU}
\\
&\hat{U}^{\dagger}_{0}(\Lambda,a) (\hat{M} - \hat{H}) \hat{U}_{0}(\Lambda,a)
= \hat{M} - \hat{H} + \frac{1}{2i} \sum_{\lambda,\lambda'} \left[ \alpha^*_{\lambda}\mathcal{U}_{\lambda\lambda'}(\Lambda,a)\hat{L}_{\lambda'} - \alpha_{\lambda}(\Lambda,a)\mathcal{U}^{\ast}_{\lambda\lambda'}(\Lambda,a)\hat{L}^{\dagger}_{\lambda'} \right] + \beta(\Lambda,a)\hat{\mathbb{I}},
\label{eq:UdgM-HU}
\end{align}
For convenience, we introduce the vectors $\hat{\bm{L}}$ and $\bm{\alpha}$ with the components $\hat{L}_{\lambda}$ and $\alpha_{\lambda}$, respectively. Then, the transformation rules 
\eqref{eq:UdgLU} and \eqref{eq:UdgM-HU} are rewritten as
\begin{align}
&\hat{U}^{\dagger}_{0}(\Lambda,a) \hat{\bm{L}} \hat{U}_{0}(\Lambda,a) = \mathcal{U}(\Lambda,a)\hat{\bm{L}} + \bm{\alpha}(\Lambda,a)\mathbb{I}, 
\label{eq:UdgLU2} 
\\
&\hat{U}^{\dagger}_{0}(\Lambda,a) (\hat{M} - \hat{H}) \hat{U}_{0}(\Lambda,a)
= \hat{M} - \hat{H} + \frac{1}{2i} \left[ \bm{\alpha}^{\dagger}(\Lambda,a)\mathcal{U}(\Lambda,a)\hat{\bm{L}} - \hat{\bm{L}}^{\dagger}\mathcal{U}^{\dagger}(\Lambda,a)\bm{\alpha}(\Lambda,a) \right] + \beta(\Lambda,a)\mathbb{I} \label{eq:UdgM-HU2}
\end{align}
Using that the operator 
$\hat{U}_{0}(\Lambda,a)$ is the unitary representation satisfying 
$\hat{U}_{0}(\Lambda'\Lambda, a' + \Lambda'a) = \hat{U}_{0}(\Lambda',a')\hat{U}_{0}(\Lambda,a)$, we can give the conditions of the unitary matrix 
$\mathcal{U}(\Lambda,a)$, the vector $\bm{\alpha}(\Lambda,a)$, and the real parameter 
$\beta(\Lambda,a)$:
\begin{align}
&\mathcal{U}(\Lambda'\Lambda, a' + \Lambda' a) = \mathcal{U}(\Lambda',a')\mathcal{U}(\Lambda,a) \label{eq:calUcalU}
\\ 
&\bm{\alpha}(\Lambda'\Lambda, a' + \Lambda' a) = \mathcal{U}(\Lambda',a')\bm{\alpha}(\Lambda,a) + \bm{\alpha}(\Lambda',a') 
\label{eq:alpha}
\\
&\beta(\Lambda'\Lambda, a' + \Lambda' a) = \beta(\Lambda,a) + \beta(\Lambda',a') + \frac{1}{2i} \left[ \bm{\alpha}^{\dagger}(\Lambda',a')\mathcal{U}(\Lambda',a')\bm{\alpha}(\Lambda,a) - \bm{\alpha}^{\dagger}(\Lambda,a)\mathcal{U}^{\dagger}(\Lambda',a)\bm{\alpha}(\Lambda',a') \right] 
\label{eq:beta}
\end{align}
It is easy to derive these conditions. 
Eqs.\eqref{eq:calUcalU} and \eqref{eq:alpha} are given by comparing the both side of 
$\hat{U}^{\dagger}_{0}(\Lambda'\Lambda, a' + \Lambda'a) \hat{\bm{L}} \hat{U}_{0}(\Lambda'\Lambda, a' + \Lambda'a) = \hat{U}^{\dagger}_{0}(\Lambda,a)\hat{U}^{\dagger}_{0}(\Lambda',a') \hat{\bm{L}} \hat{U}_{0}(\Lambda',a')\hat{U}_{0}(\Lambda,a)$. 
Eq.\eqref{eq:beta} is obtained by comparing the both side of $\hat{U}^{\dagger}_{0}(\Lambda'\Lambda, a' + \Lambda'a)(\hat{M}-\hat{H})\hat{U}_{0}(\Lambda'\Lambda, a' + \Lambda'a) =  \hat{U}^{\dagger}_{0}(\Lambda,a)\hat{U}^{\dagger}_{0}(\Lambda',a') (\hat{M}-\hat{H})\hat{U}_{0}(\Lambda',a')\hat{U}_{0}(\Lambda,a)$.

Our task is to derive the expression of the self-adjoint operator 
$\hat{M} - \hat{H}$ and the Lindblad operator 
$\hat{\bm{L}}$ satisfying Eq.\eqref{eq:UdgLU2} and \eqref{eq:UdgM-HU2}. 
In particular, Eq.\eqref{eq:UdgLU2} is decomposed into irreducible representation subspaces for ease of conputation. 
Hence, the irreducible unitary representations of the Poincar\'{e} group are useful for our analysis. 
We introduce the standard momentum $\ell^\mu$ and the Lorentz transformation $\left( S_{q} \right)^{\mu}_{\ \nu}$ with 
\begin{equation}
q^{\mu} = \left( S_{q} \right)^{\mu}_{\ \nu} \ell^\nu. \label{eq:q}
\end{equation}
From \eqref{eq:calUcalU}, the unitary matrix 
$\mathcal{U}(\Lambda,a)$ is written as
\begin{equation}
    \mathcal{U}(\Lambda,a) = \mathcal{U}(I,a)\mathcal{U}(\Lambda,0) = \mathcal{T}(a)\mathcal{V}(\Lambda), \label{eq:U(Lambda,a)},
\end{equation}
where 
$I$ is the identity matrix, 
$\mathcal{U}(I,a) = \mathcal{T}(a)$, $\mathcal{U}(\Lambda,0) = \mathcal{V}(\Lambda)$. 
The unitary matrix $\mathcal{U}(\Lambda,a)$ is the unitary representation on a vector space, and then we define the following vector on the vector space:
\begin{equation}
    R^{\mu}\bm{v}_{\ell,\xi} = \ell^\mu\bm{v}_{\ell,\xi}, \label{eq:R}
\end{equation}
where 
$R^{\mu}$ is the generator of spacetime translation, the label $\xi$ describes the degrees of freedom which can not be specified by momentum 
$\ell^\mu$.
Also, $\mathcal{T}(a)$ can be written as $e^{-i R_{\mu}a^{\mu}}$ by using the generator of spacetime translation $R^{\mu}$.
Now, we define the eigenvector $\bm{v}_{q,\xi}$ which belongs to the eigenvalue 
$q^{\mu}$ of 
$R^{\mu}$ as
\begin{equation}
\bm{v}_{q,\xi} = N_{q} \mathcal{V}(S_{q})\bm{v}_{\ell,\xi}, 
\label{eq:vq}
\end{equation}
where $N_{q}$ is the normalization. Then, we can obtain the following rules 
$\bm{v}_{q,\xi}$:
\begin{align}
&\mathcal{T}(a) \bm{v}_{q,\xi} = e^{-iq^{\mu}a_{\mu}}\bm{v}_{q,\xi}, 
\label{eq:T} 
\\
&\mathcal{V}(\Lambda)\bm{v}_{q,\xi} = \frac{ N_{q} }{ N_{\Lambda q} } \sum_{\xi'} \mathcal{D}_{\xi\xi'} \left( W(\Lambda,q) \right)\bm{v}_{\Lambda q,\xi'}, 
\label{eq:V}
\end{align}
where 
$W(\Lambda,q) = S^{-1}_{\Lambda q}\Lambda S_{q}$ is an element of the little group and satisfies 
$W^{\mu}_{\ \nu}\ell^\nu = \ell^\mu$. 
For the derivations of Eqs.\eqref{eq:T} and \eqref{eq:V}, see Ref.\cite{Weinberg1995}.
The matrix 
$\mathcal{D}(W)$ with the components
$\mathcal{D}_{\xi\xi'}(W)$ forms the unitary representation of the little group. 
For the various four-momentum, we give the standard momenta and the little groups corresponded to each standard momentum in Table \ref{tab:1}. 
For simplicity, $\xi$ is assumed to be the label of basis vectors of the irreducible representation subspaces of the little group.
\begin{table}[H]
\centering
\begin{tabular}{|c|c|}
\hline
Standard Momentum $\ell^\mu$ & Little Group 
\\ \hline
$\ell^\mu = [ M, 0, 0, 0], \ M > 0$  & SO(3) 
\\ \hline
$\ell^\mu = [ -M, 0, 0, 0], \ M > 0$ & SO(3) 
\\ \hline
$\ell^\mu = [ \kappa, 0, 0, \kappa], \ \kappa > 0 $ & ISO(2) 
\\ \hline
$\ell^\mu = [ -\kappa, 0, 0, \kappa], \ \kappa > 0 $ & ISO(2) 
\\ \hline
$\ell^\mu = [ 0, 0, 0, N], \ N^{2} > 0$ & SO(2,1) 
\\ \hline
$\ell^\mu = [0, 0, 0, 0] $ & SO(3,1) 
\\ \hline
\end{tabular}
\caption{Classification of the standard momentum $\ell^\mu$ and the little group associated with $\ell^\mu$, $M$ is mass and $\kappa$ is arbitrary positive energy.}
\label{tab:1}
\end{table}
Next, we consider the unitary operator $\hat{U}_{0}(\Lambda,a)$. 
As in the unitary matrix $\mathcal{U}(\Lambda,a)$, 
the operator $\hat{U}_{0}(\Lambda,a)$ is decomposed in the same way. 
Therefore, it is written as
\begin{equation}
\hat{U}_{0}(\Lambda,a) = \hat{U}_{0}(I,a)\hat{U}_{0}(\Lambda,0) = \hat{T}(a)\hat{V}(\Lambda), 
\label{eq:U0}
\end{equation}
where 
$\hat{U}_{0}(I,a) = \hat{T}(a) = e^{-i\hat{P}_{\mu}a^{\mu}}$ with $\hat{P}^{\mu} = [ \hat{H}, \hat{P}^{1}, \hat{P}^{2},\hat{P}^{3} ]$ and $\hat{U}_{0}(\Lambda,0)=\hat{V}(\Lambda)$ with the generators $\hat{J}^{i}$ and $\hat{K}^{i}_{0}$. 
Let us focus on the transformation rule Eq.\eqref{eq:UdgLU2}. 
From the rule for 
$\Lambda = I$, we obtain
\begin{equation}
\hat{T}^{\dagger}(a)\hat{\bm{L}}\hat{T}(a) = \mathcal{T}(a)\hat{\bm{L}} + \bm{\alpha}(I,a)\hat{\mathbb{I}}. 
\label{eq:TdgLT}
\end{equation}
In Eq. \eqref{eq:UdgLU2} for 
$a^{\mu} = 0$, we obtain
\begin{equation}
\hat{V}^{\dagger}(\Lambda)\hat{\bm{L}}\hat{V}(\Lambda) = \mathcal{V}(\Lambda)\hat{\bm{L}} + \bm{\alpha}(\Lambda,0)\hat{\mathbb{I}}. \label{eq:VdgLV}
\end{equation}
Introducing 
$\hat{L}_{q,\xi} =\bm{v}_{q,\xi}^{\dagger}\hat{\bm{L}}$ and 
$\alpha_{q,\xi} = \bm{v}_{q,\xi}^{\dagger} \bm{\alpha}$, w
e can rewrite \eqref{eq:TdgLT} and \eqref{eq:VdgLV} as follows:
\begin{align}
\hat{T}^{\dagger}(a)\hat{L}_{q,\xi}\hat{T}(a) &= e^{-iq_{\mu}a^{\mu}}\hat{L}_{q,\xi} + \alpha_{q,\xi}(I,a)\hat{\mathbb{I}},
\label{eq:TdgLT2}
\\
\hat{V}^{\dagger}(\Lambda)\hat{L}_{q,\xi}\hat{V}(\Lambda)
&= \frac{N^{\ast}_{q}}{N^{\ast}_{\Lambda^{-1}q}} \sum_{\xi'} \mathcal{D}^{\ast}_{\xi'\xi}\left( W(\Lambda^{-1},q) \right) \hat{L}_{\Lambda^{-1}q,\xi'} + \alpha_{q,\xi}(\Lambda,0)\hat{\mathbb{I}} \label{eq:VdgLV2}.
\end{align}
Because of these transformation rules, we get the expression of the complex number $\alpha_{q,\xi}(\Lambda,a)$. 
Since \eqref{eq:UdgLU2} can be rewritten as
\begin{equation}
\hat{U}^{\dagger}_{0}(\Lambda,a)\hat{L}_{q,\xi}\hat{U}_{0}(\Lambda,a) = \bm{v}^{\dagger}_{q,\xi}\mathcal{U}(\Lambda,a)\hat{\bm{L}} + \alpha_{q,\xi}(\Lambda,a)\hat{\mathbb{I}}, 
\label{eq:UdgLU3}
\end{equation}
we get 
$\alpha_{q,\xi}(\Lambda,a)$ as
\begin{equation}
\alpha_{q,\xi}(\Lambda,a) = e^{-iq_{\mu}a^{\mu}}\alpha_{q,\xi}(\Lambda,0) + \alpha_{q,\xi}(I,a)
\label{eq:alpha2}
\end{equation}
by using \eqref{eq:TdgLT2} and \eqref{eq:VdgLV2} and comparing the both side of Eq.\eqref{eq:UdgLU3}.
In Eq.\eqref{eq:VdgLV2} for 
$\Lambda = S_{q}$, we obtain
\begin{equation}
\hat{V}^{\dagger}(S_{q})\hat{L}_{q,\xi}\hat{V}(S_{q}) = N^{\ast}_{q}\hat{L}_{\ell,\xi} + \alpha_{q,\xi}(S_{q},0)\hat{\mathbb{I}}, 
\label{eq:VSL}
\end{equation}
where we used the facts that 
$N_\ell=1$ and 
$W\left( S^{-1}_{q},q \right) = S^{-1}_{S^{-1}_{q}q}S^{-1}_{q}S_{q} = S^{-1}_\ell = I$ hold. 
These facts can be checked by using Eq.\eqref{eq:vq} and the definition of $W(\Lambda,q)$. 
Eq.\eqref{eq:VSL} means that the Lindblad operator 
$\hat{L}_{q,\xi}$ is determined from the Lindblad operator 
$\hat{L}_{\ell,\xi}$ with the standard momentum $\ell^\mu$. To discuss the form of the Lindblad operator $\hat{L}_{\ell,\xi}$, we focus on the following equations obtained from Eqs.\eqref{eq:TdgLT2} and \eqref{eq:VdgLV2} for 
$q^{\mu} = \ell^\mu$ and 
$\Lambda = Q$ with $Q^{\mu}_{\ \nu}\ell^\nu = \ell^\mu$, respectively:
\begin{align}
&\hat{T}^{\dagger}(a)\hat{L}_{\ell,\xi}\hat{T}(a) = e^{-il_{\mu}a^{\mu}}\hat{L}_{\ell,\xi} + \alpha_{\ell,\xi}(I,a)\hat{\mathbb{I}}, 
\label{eq:TdgLlT}
\\
&\hat{V}^{\dagger}(Q)\hat{L}_{\ell,\xi}\hat{V}(Q) = \sum_{\xi'}\mathcal{D}^{\ast}_{\xi'\xi}(Q^{-1})\hat{L}_{\ell,\xi'} + \alpha_{\ell,\xi}(Q,0)\hat{\mathbb{I}}, \label{eq:VdgLlV}
\end{align}
where note that 
$N_\ell = N_{Q^{-1}\ell} = 1$. 

Now, we can get a model of the Markovian QME for a massive spin-0 particle by assuming that the Lindblad operator $\hat{L}_{q,\xi}$ is given as 
\begin{equation}
\hat{L}_{q,\xi} = \int d^{3}p \ f_{q,\xi}(\bm{p})\hat{a}(\bm{p}), 
\label{eq:Lansatz}
\end{equation}
and that the self-adjoint operator $\hat{M} - \hat{H}$ is given as
\begin{equation}
\hat{M} - \hat{H} = \int d^{3}p \ d^{3}p' \   g(\bm{p},\bm{p}')\hat{a}^{\dagger}(\bm{p})\hat{a}(\bm{p}'), 
\label{eq:Mansatz}
\end{equation}
where 
$g(\bm{p},\bm{p}') = g^*(\bm{p}',\bm{p})$ is satisfied because the 
$\hat{M} - \hat{H}$ is self-adjoint. 
The Markovian QME with the above operators gives an evolution from a Gausssian state to another Gausssian state. 
Furthermore, in the dynamics, there are no particle creations because the GKSL generator 
$\mathcal{L}$ has no creation processes.
The Poincar\'{e} transformation rules of $\hat{a}^{\dagger}(\bm{p})$ shown in Ref.\cite{Peres2004,Weinberg1995} 
\begin{align}
&\hat{T}(a)\hat{a}^{\dagger}(\bm{p})\hat{T}^{\dagger}(a) = e^{-ip^{\mu}a_{\mu}}\hat{a}^{\dagger}(\bm{p}), 
\label{eq:TaTdg} 
\\
&\hat{V}(\Lambda)\hat{a}^{\dagger}(\bm{p})\hat{V}^{\dagger}(\Lambda) = \sqrt{\frac{E_{\bm{p}_{\Lambda}}}{E_{\bm{p}}}} \  \hat{a}^{\dagger}(\bm{p}_{\Lambda}), 
\label{eq:VaVdg}
\end{align}
are useful to obtain the model of the massive particle, where 
$E_{\bm{p}} = p^{0}$, $ E_{\bm{p}_{\Lambda}} = (\Lambda p)^{0} $ and $(\bm{p}_{\Lambda})^{i}$ is the vector with the elements written as $(\bm{p}_{\Lambda})^{i} = (\Lambda p)^{i}$ \cite{Weinberg1995}. 
Also, $W(\Lambda,p) = S^{-1}_{\Lambda p}\Lambda S_{p}$ with $(S_{p})^{\mu}_{\ \nu}k^{\nu} = p^{\mu}$ is an element of the little group and satisfies $W^{\mu}_{\ \nu} k^{\nu} = k^{\mu}$, where $k^{\mu}$ is the standard momentum for a massive particle ( $k^{\mu} = [ m,0,0,0 ], m > 0$ ).
Substituting the ansatz of $\hat{L}_{q,\xi}$ \eqref{eq:Lansatz} into Eqs.\eqref{eq:TdgLlT} and \eqref{eq:VdgLlV} and using the transformation rules \eqref{eq:TaTdg} and \eqref{eq:VaVdg}, we obtain the equations
\begin{align}
f_{\ell,\xi}(\bm{p})e^{-ip^{\mu}a_{\mu}} &= f_{\ell,\xi}(\bm{p})e^{-i\ell^\mu a_\mu}, 
\label{eq:Tf}
\\
\sqrt{\frac{E_{\bm{p}_{Q}}}{E_{\bm{p}}}} f_{\ell,\xi}(\bm{p}_{Q}) &= \sum_{\xi'}\mathcal{D}^*_{\xi'\xi}(Q^{-1})f_{\ell,\xi'}(\bm{p}). 
\label{eq:Vf}
\end{align}
In addition, we get 
$\alpha_{q,\xi}(\Lambda,0) = \alpha_{q,\xi}(I,a)=0$, and then Eq.\eqref{eq:alpha2} leads to 
$\alpha_{q,\xi}(\Lambda,a)=0$. 
Also, we can obtain the condition of the coefficient 
$g(\bm{p},\bm{p}')$ by substituting the ansatz of 
$\hat{M} - \hat{H}$ \eqref{eq:Mansatz} into Eq.\eqref{eq:UdgM-HU} as
\begin{equation}
\sqrt{ \frac{ E_{\bm{p}_{\Lambda}}E_{\bm{p}'_{\Lambda}} }{ E_{\bm{p}}E_{\bm{p}'} } } \ g(\bm{p}_{\Lambda},\bm{p}'_{\Lambda})e^{i\left( (\Lambda p)^{\mu} - (\Lambda p')^{\mu} \right)a_{\mu}} = g(\bm{p},\bm{p}'). 
\label{eq:g}
\end{equation}
From the fact that there is no term proportional to the identity operator $\hat{\mathbb{I}}$ in 
$\hat{M} - \hat{H}$ and the condition $\alpha_{q,\xi}(\Lambda,a) = 0$, the real number 
$\beta(\Lambda,a) $ vanishes.

Solving Eqs.\eqref{eq:Tf}, \eqref{eq:Vf}, and \eqref{eq:g}, we can derive the Markovian QME for a massive spin-0 particle as follows:
\begin{equation}
\frac{d}{dt} \rho(t) = -i \left[ \hat{H} + g\hat{N}, \rho(t) \right] + \gamma \int d^{3}p \left[ \hat{a}(\bm{p})\rho(t)\hat{a}^{\dagger}(\bm{p}) - \frac{1}{2} \left\{ \hat{a}^{\dagger}(\bm{p})\hat{a}(\bm{p}), \rho(t) \right\} \right], 
\label{eq:massive}
\end{equation}
where a non-negative parameter 
$\gamma$, a real parameter $g$, and $\hat{N}$ is the number operator \eqref{eq:N}. 
The Markovian QME, Eq.\eqref{eq:model} presented in Sec.\ref{sec:model}, is Eq.\eqref{eq:massive} for 
$g=0$. 
The detailed derivation of Eq.\eqref{eq:massive} is presented in Appendix \ref{app:massive}.

\section{Discussion}

In this section, we discuss the previous works on relativistic Markovian QME \cite{Poulin2017, Alicki1986, Diosi2022} in comparing it with the presented Markovian QME.
In \cite{Poulin2017, Alicki1986}, the following master equation of a massive spin-0 particle is proposed:
\begin{equation}
\frac{d}{dt} \rho(t)=\mathcal{H}[\rho(t)]+\tilde{\mathcal{D}}[\rho(t)],
\label{eq:Poulin}
\end{equation}
where 
$\mathcal{H}[\rho(t)]=-i[\hat{H},\rho(t)]$ with the Hamiltonian $\hat{H}$ \eqref{eq:H}, and 
\begin{equation}
\tilde{\mathcal{D}}[\rho(t)]=\kappa \int d^{3}p E_{\bm{p}} \left[ \hat{a}(\bm{p})\rho(t)\hat{a}^{\dagger}(\bm{p}) - \frac{1}{2} \left\{ \hat{a}^{\dagger}(\bm{p})\hat{a}(\bm{p}), \rho(t) \right\} \right].
\label{eq:tildeD}
\end{equation}
The dissipation rate associated with the dissipator 
$\tilde{\mathcal{D}}$ is read out as 
$\kappa E_{\bm{p}}$, which depends on the energy of the particle. 
The master equation \eqref{eq:Poulin} was derived by respecting the Poincar\'{e} (or Lorentz) covariance \cite{Poulin2017, Alicki1986}. 
From this master equation \eqref{eq:Poulin}, we have the field equation of evolution, 
\begin{equation}
\frac{\partial^2}{\partial t^2}\hat{\varphi}_\text{H}(t,\bm{x})+\kappa \sqrt{-\nabla^2+m^2} \frac{\partial}{\partial t}  \hat{\varphi}_\text{H} (t,\bm{x})=\left(1+\frac{\kappa^2}{4} \right)(\nabla^2-m^2) \hat{\varphi}_\text{H}(t,\bm{x}),
\label{eq:varphiH}
\end{equation}
where 
$\hat{\varphi}_\text{H}(t,\bm{x})=e^{\tilde{\mathcal{L}}^\dagger t}[\Phi(\bm{x})]$ with the adjoint $\tilde{\mathcal{L}}^\dagger$ of
$\tilde{\mathcal{L}}=\mathcal{H}+\tilde{\mathcal{D}}$ and the initial condition 
$\hat{\Phi}(\bm{x})$ given in \eqref{eq:Phi}. 
We also find that the field 
$\hat{\varphi}_\text{H}(t,\bm{x})$ does not satisfy the microcausality, that is  
$[\hat{\varphi}_\text{H}(x^0,\bm{x}), \hat{\varphi}_\text{H}(y^0,\bm{y})]\neq 0$ even for 
$(x^0-y^0)^2-(\bm{x}-\bm{y})^2<0$, and  the violation of microcausality typically appears in the range 
$1/m$ \cite{Poulin2017, Diosi2022}. 
In contrast, our QME is based on not the Poincar\'{e} covariance but the invariance defined by \eqref{eq:Pinv}. 
This results in the dissipator $\mathcal{D}$ in our QME, which gives the dissipation rate 
independent of the energy of the particle.
Furthermore, as we checked, the microcausality holds in our model. 

It is interesting to highlight the difference between the two dissipators 
$\mathcal{D}$ and 
$\tilde{\mathcal{D}}$ by following the discussion in Ref.\cite{Diosi2022}.
We pick up two critiques given in \cite{Diosi2022} on the above master equation \eqref{eq:Poulin}. 
One is that the mater equation \eqref{eq:Poulin} does not fully respect the Lorentz symmetry. 
Concretely, this means that the dissipator 
$\tilde{\mathcal{D}}$ is not invariant under the Lorentz boost. 
On the contrary, our dissipator 
$\mathcal{D}$ in \eqref{eq:model} is invariant under the Lorentz boost since 
$\mathcal{D}=\mathcal{L}-\mathcal{H}$ satisfying \eqref{eq:L-H} is invariant under any Poincar\'{e} transformations including the Lorentz boost. 

Another critique is that the field of the massive particle does not follow the covariant equation of evolution. 
Indeed, it seems that not only the above field equation \eqref{eq:varphiH} but also our field equation \eqref{eq:diss.KG} are not Lorentz covariant.
However, we guess that our field is not scalar and this fact makes its equation covariant. 
To clarify this argument, it may be important to note that the field transformation in quantum theory is identified through unitary transformations. 
The scalar field of a massive spin-0 particle, 
$\phi(x)$, changes as  
$\phi(x) \rightarrow \phi'(x')=\phi(x)$ under Lorentz transformations 
$x'^\mu=\Lambda^\mu{}_\nu x^\nu$. 
In unitary relativistic quantum theories, such as quantum field theories, the corresponding transformation is 
\begin{equation}
\hat{\phi}_\text{H}(x) \rightarrow \hat{V}^\dagger(\Lambda) \hat{\phi}_\text{H}(x') \hat{V}(\Lambda)=\hat{\phi}_\text{H}(x),
\label{eq:phiH}
\end{equation}
where the unitary operator
$\hat{V}(\Lambda)=\hat{U}_0 (\Lambda,0)$ defined around \eqref{eq:U0} represents the Lorentz transformation, and 
$\hat{\phi}_\text{H}(t,\bm{x})=e^{i\hat{H}t}\hat{\Phi}(\bm{x})e^{-i\hat{H}t}$. 
On the other hand, the field operator 
$\hat{\Phi}_\text{H}$ obeying the dissipative KG equation \eqref{eq:diss.KG} in our model is not transformed like \eqref{eq:phiH}. 
The field
$\hat{\Phi}_\text{H}$ satisfies 
$\hat{V}^\dagger(\Lambda) e^{\frac{\gamma}{2}x'^0} \hat{\Phi}_\text{H}(x') \hat{V}(\Lambda)=e^{\frac{\gamma}{2}x^0} \hat{\Phi}_\text{H}(x)$ because 
$\hat{\Phi}_\text{I}(x)=e^{\frac{\gamma}{2}x^0} \hat{\Phi}_\text{H}(x)$ is the solution of usual free KG equation, which plays a role of Lorentz scalar field. 
Hence the field transformation of 
$\hat{\Phi}_\text{H}$ is given as 
\begin{equation}
\hat{\Phi}_\text{H}(x) \rightarrow \hat{V}^\dagger (\Lambda)\hat{\Phi}_\text{H}(x')\hat{V}(\Lambda)=e^{-\gamma(x'^{0}-x^0)/2} \hat{\Phi}_\text{H}(x).
\label{eq:PhiH.trans}
\end{equation}
This means that 
$\hat{\Phi}_\text{H}(x)$ is not Lorentz scalar field, which is crucial to discuss the Lorentz covariance of \eqref{eq:diss.KG}. 
If the field follows the transformation \eqref{eq:PhiH.trans}, the dissipative KG equation \eqref{eq:diss.KG} is certainly Lorentz covariant, which is easily observed from the fact that $\hat{\Phi}_\text{I}(x)$ is just a Lorentz scalar field satisfying the free KG equation.
We think that the present discussion on the covariance of the field equation is very speculative. 
We would need a clear understanding on this concern as a future issue.


\section{Conclusion and outlook}

In this study, we discussed the Markovian QME in the GKSL form, whose solution is given by a quantum dynamical semigroup with the Poincar\'{e} invariance. 
In particular, we derived the Markovian QME of a relativistic massive spin-0 particle and investigated its field dynamics. 
First, it turned out that the field of the massive particle follows the KG equation with a dissipation term. 
Besides, we also showed that not only field operators but also any local operators commute with each other when these operators are space-like separated. 
This means that the present theory specified on the Markovian QME has the microcausality property. 
Our formulation for the Markovian QME respecting the Poincar\'{e} symmetry may give dissipative quantum field theories. 

In this paper, we just provided a theoretical framework for yielding the Markovian QME of relativistic particles.
We expect that this framework may be valid for relativistic dissipative phenomena, such as the decay of unstable relativistic particles. 
Confirming this expectation is a future subject of this present paper, which would deepen the understanding of our framework.

Another future subject is to build a theoretical framework including the two standpoints mentioned in the Introducition: one is that ``gravity is quantum", and the other is that ``gravity is classical". 
For this purpose, we will expand the present theory to that can take into account gravitational interactions. 
If such a framework is established, we can provide candidate theories that describe the interplay regime of quantum and gravity phenomena. 
This leads to an interesting theme of what candidates will be accepted from future quantum gravity experiments. 
In particular, we believe that recent experimental technologies\cite{Cripe2019, Matsumoto2020, Westphal2021} for testing quantum mechanics and gravity theory would explore the consistent theory unifying quantum and gravity physics. 
We hope that the present work will be a help for the research of quantum gravity.

\begin{acknowledgements}
We thank D. Carney, K. Gallock-Yoshimura, B. L. Hu, S. Kanno, Y. Kuramochi, Y. Nambu, J. Oppenheim, and K. Yamamoto for variable discussions and comments related to this paper. 
A.M. was supported by JSPS KAKENHI (Grants No.~JP23K13103 and No.~JP23H01175).
\end{acknowledgements}

\begin{appendix}

\section{Proof that the GKSL generator in Eq.\eqref{eq:model} is Poincar\'{e} invariant}
\label{app:model}

We prove the Poincar\'{e} invariance of the GKSL generator  
$\mathcal{L}$ in Eq.\eqref{eq:model}. 
Since the GKSL generator
$\mathcal{L}$ has the form 
$\mathcal{L}=\mathcal{H}+\mathcal{D}$, the condition of Poincar\'{e} invariance \eqref{eq:L-H} is nothing but the invariance condition for the dissipator $\mathcal{D}$, 
\begin{equation}
\hat{U}_{0}(\Lambda,a)\mathcal{D}\left[ \rho(0) \right] \hat{U}^{\dagger}_{0}(\Lambda,a) = \mathcal{D} \left[ \hat{U}_{0}(\Lambda,a)\rho(0)\hat{U}^{\dagger}_{0}(\Lambda,a) \right],
\label{eq:DinvA} 
\end{equation}
where 
$\mathcal{D}$ is yielded as
\begin{equation}
\mathcal{D}[\rho(0)]= 
\gamma \int d^{3}p \left[ \hat{a}(\bm{p})\rho(0)\hat{a}^{\dagger}(\bm{p}) - \frac{1}{2} \left\{ \hat{a}^{\dagger}(\bm{p})\hat{a}(\bm{p}), \rho(0) \right\} \right].
\label{eq:DA}
\end{equation}
Substituting 
$\mathcal{D}[\rho(0)]$ into the left hand side of the condition \eqref{eq:DinvA}, the left hand side can be rewritten as follow:
\begin{align}
&\hat{U}_{0}(\Lambda,a)\mathcal{D}\left[ \rho(0) \right] \hat{U}^{\dagger}_{0}(\Lambda,a)
\nonumber 
\\
&
\quad 
=\gamma \int d^{3}p \ \hat{U}_{0}(\Lambda,a) \left[ \hat{a}(\bm{p})\rho(0)\hat{a}^{\dagger}(\bm{p}) - \frac{1}{2} \left\{ \hat{a}^{\dagger}(\bm{p})\hat{a}(\bm{p}), \rho(0) \right\} \right] \hat{U}^{\dagger}_{0}(\Lambda,a) 
\nonumber 
\\
&
\quad
=\gamma \int d^{3}p \left[ \hat{U}_{0}(\Lambda,a) \hat{a}(\bm{p})\rho(0)\hat{a}^{\dagger}(\bm{p})\hat{U}^{\dagger}_{0}(\Lambda,a) - \frac{1}{2} \hat{U}_{0}(\Lambda,a) \left\{ \hat{a}^{\dagger}(\bm{p})\hat{a}(\bm{p}), \rho(0) \right\}\hat{U}^{\dagger}_{0}(\Lambda,a) \right] 
\nonumber 
\\
&
\quad
=\gamma \int d^{3}p \ \hat{U}_{0}(\Lambda,a) \hat{a}(\bm{p}) \hat{U}^{\dagger}_{0}(\Lambda,a) \left[ \hat{U}_{0}(\Lambda,a) \rho(0) \hat{U}^{\dagger}_{0}(\Lambda,a) \right] \hat{U}_{0}(\Lambda,a) \hat{a}^{\dagger}(\bm{p})\hat{U}^{\dagger}_{0}(\Lambda,a) 
\nonumber 
\\
&
\quad \quad 
-\frac{\gamma}{2} \int d^{3}p \left\{ \hat{U}_{0}(\Lambda,a) \hat{a}^{\dagger}(\bm{p})\hat{U}^{\dagger}_{0}(\Lambda,a)\hat{U}_{0}(\Lambda,a) \hat{a}(\bm{p})\hat{U}^{\dagger}_{0}(\Lambda,a), \hat{U}_{0}(\Lambda,a)\rho(0)\hat{U}^{\dagger}_{0}(\Lambda,a) \right\}. 
\label{eq:left.D}
\end{align}
Using the transformation rules 
\eqref{eq:TaTdg} and \eqref{eq:VaVdg}, which lead to
\begin{equation}
\hat{U}_{0}(\Lambda,a)\hat{a}^{\dagger}(\bm{p})\hat{U}^{\dagger}_{0}(\Lambda,a)
= \sqrt{\frac{ E_{\bm{p}_{\Lambda}} }{E_{\bm{p}}}} e^{-i(\Lambda p)^\mu a_\mu} \hat{a}^{\dagger}(\bm{p}_{\Lambda}), \label{eq:UaUdg}
\end{equation}
we have
\begin{align}
&\hat{U}_{0}(\Lambda,a) \mathcal{D} \left[ \rho(0) \right] \hat{U}^{\dagger}_{0}(\Lambda,a)
\nonumber
\\
&
\quad 
=\gamma \int d^{3}p \ \hat{U}_{0}(\Lambda,a) \hat{a}(\bm{p}) \hat{U}^{\dagger}_{0}(\Lambda,a) \left[ \hat{U}_{0}(\Lambda,a) \rho(0) \hat{U}^{\dagger}_{0}(\Lambda,a) \right] \hat{U}_{0}(\Lambda,a) \hat{a}^{\dagger}(\bm{p})\hat{U}^{\dagger}_{0}(\Lambda,a) 
\nonumber 
\\
& 
\quad \quad 
- \frac{\gamma}{2} \int d^{3}p \left\{ \hat{U}_{0}(\Lambda,a) \hat{a}^{\dagger}(\bm{p})\hat{U}^{\dagger}_{0}(\Lambda,a)\hat{U}_{0}(\Lambda,a) \hat{a}(\bm{p})\hat{U}^{\dagger}_{0}(\Lambda,a), \hat{U}_{0}(\Lambda,a)\rho(0)\hat{U}^{\dagger}_{0}(\Lambda,a) \right\}
\nonumber 
\\
&
\quad 
= \gamma \int d^{3}p \ \frac{ E_{\bm{p}_{\Lambda}} }{E_{\bm{p}}} \left[ \hat{a}(\bm{p}_{\Lambda}) \hat{U}_{0}(\Lambda,a) \rho(0) \hat{U}^{\dagger}_{0}(\Lambda,a) \hat{a}^{\dagger}(\bm{p}_{\Lambda}) - \frac{1}{2} \left\{ \hat{a}^{\dagger}(\bm{p}_{\Lambda})\hat{a}(\bm{p}_{\Lambda}), \hat{U}_{0}(\Lambda,a) \rho(0) \hat{U}^{\dagger}_{0}(\Lambda,a) \right\} \right] 
\nonumber 
\\
&
\quad 
= \gamma \int d^{3}p_{\Lambda} \left[ \hat{a}(\bm{p}_{\Lambda}) \hat{U}_{0}(\Lambda,a) \rho(0) \hat{U}^{\dagger}_{0}(\Lambda,a) \hat{a}^{\dagger}(\bm{p}_{\Lambda}) - \frac{1}{2} \left\{ \hat{a}^{\dagger}(\bm{p}_{\Lambda})\hat{a}(\bm{p}_{\Lambda}), \hat{U}_{0}(\Lambda,a) \rho(0) \hat{U}^{\dagger}_{0}(\Lambda,a) \right\} \right]
\nonumber 
\\
&
\quad 
= \mathcal{D} \left[ \hat{U}_{0}(\Lambda,a) \rho(0) \hat{U}^{\dagger}_{0}(\Lambda,a) \right],
\label{eq:DinvA2}
\end{align}
where we used the fact that 
$d^3p/E_\text{p}$ is the Lorentz ionvariant measure in the third equality. 
Certainly, it is confirmed that the GKSL generator
$\mathcal{L}$ of the Markovian quantum master equation \eqref{eq:model} satisfies the condition Eq.\eqref{eq:L-H}.

\section{Proof of the commutation relations Eqs.\eqref{eq:PhiPhi} and and \eqref{eq:PiPhi}}
\label{sec:pf.comm.}

Before the proofs of them, we mention the Lorentz invariance of the measure 
$d^{3}p/2E_{\bm{p}} $. 
This measure can be rewirtten as follow:
\begin{equation}
\frac{d^{3}p}{ 2E_{\bm{p}} } = d^{4}p \ \delta(p_{\mu}p^{\mu} + m^{2})\theta(p^{0}), 
\label{eq:inv.meas.} 
\end{equation}
where 
$\theta(p^{0})$ is step function defined as 
\begin{equation}
\theta(p^{0}) =
\left\{ \,
\begin{aligned}
 1 \ (p^{0} \geq 0) \\
 0 \ (p^{0} < 0)
    \end{aligned}
    \right.,
\label{eq:step}
\end{equation}
and 
$\eta_{\mu\nu}$ is the Minkowski metric tensor. 
Under the Lorentz transformation 
$p^\mu \rightarrow p'^\mu=\Lambda^\mu{}_\nu p^\nu$, 
the right hand side of \eqref{eq:inv.meas.} is
\begin{align}
d^{4}p' \, \delta(p'_{\mu}p'^{\mu} + m^{2})\theta(p'^{0}) 
&= d^{4}p \ \delta(  p_\mu p^\mu + m^{2} )\theta(p'^{0}), 
\label{eq:inv.meas.2}
\end{align} 
where the matrix 
$\Lambda^{\mu}_{\ \nu}$ is a proper orthochronous Lorentz transformation matrix. 
If 
$p'^{0}$ is positive, the step function is Lorentz invariant too. 
Looking the delta function, the integration have a meaningful value when 
$p^{\rho}p_{\rho} + m^{2} = 0$. 
The condition 
$p^{\rho}p_{\rho} + m^{2} = 0$ means that 
$p^{\rho}$ is time-like vector. 
Since we consider a proper orthochronous Lorentz transformation, 
$\theta(p^{0}) = \theta(p'^{0}) = 1$ holds. 
Thus, we obtain the fact that the measure 
$d^{3}p/2E_{\bm{p}}$ is Lorentz invariant. 

We first prove Eq.\eqref{eq:PhiPhi}. The commutator 
$\left[ \hat{\Phi}_\text{H}(x^{0},\bm{x}), \hat{\Phi}_\text{H}(y^{0},\bm{y}) \right]$ is computed as
 \begin{align}
\left[ \hat{\Phi}_\text{H}(x^{0},\bm{x}), \hat{\Phi}_\text{H}(y^{0},\bm{y}) \right]
&= e^{ -\frac{\gamma}{2}( x^{0} + y^{0} ) } \left[ \hat{\Phi}_\text{I}(x^{0},\bm{x}), \hat{\Phi}_\text{I}(y^{0},\bm{y}) \right] 
\nonumber
\\
&= \frac{e^{-\frac{\gamma}{2}(x^{0} + y^{0})}}{ (2\pi)^{3} }  \int \frac{d^{3}p}{\sqrt{2E_{\bm{p}}}} \int \frac{d^{3}p'}{\sqrt{2E_{\bm{p}'}}}
\nonumber 
\\  
& 
\quad 
\times \left[ e^{i(p^{\mu}x_{\mu} - p'^{\mu}y_{\mu}) } \left[ \hat{a}(\bm{p}), \hat{a}^{\dagger}(\bm{p}') \right] - e^{-i(p^{\mu}x_{\mu} - p'^{\mu}y_{\mu}) } \left[ \hat{a}(\bm{p}), \hat{a}^{\dagger}(\bm{p}') \right] \right]
\nonumber 
\\
&= \frac{e^{-\frac{\gamma}{2}(x^{0} + y^{0})}}{ (2\pi)^{3} } \int \frac{d^{3}p}{2E_{\bm{p}}} \left[ e^{ip^{\mu}(x_{\mu} - y_{\mu})} - e^{-ip^{\mu}(x_{\mu} - y_{\mu})} \right],
\label{eq:comPhi}
\end{align}
where we used the notations 
$p^{\mu} = [E_{\bm{p}},\bm{p}]$, 
$x_{\mu} = [-x^{0},\bm{x}]$, 
$y_{\mu} = [-y^{0},\bm{y}]$, and Eqs.\eqref{eq:aadg}. 
For any spacelike distance 
$(x^0-y^0)^2-(\bm{x}-\bm{y})^2<0$, we can introduce 
$x'_{\rho} - y'_{\rho} = \Lambda^{\mu}_{\ \rho}(x_{\mu} - y_{\mu})$ and 
choose a proper orthochronous Lorentz transformation $\Lambda$ such that $x'_{0} - y'_{0} = 0$.
Thus, we obtain the following result
\begin{align}
\left[ \hat{\Phi}_\text{H}(x^{0},\bm{x}), \hat{\Phi}_\text{H}(y^{0},\bm{y}) \right]
&= \frac{1}{ (2\pi)^{3} } e^{-\frac{\gamma}{2}(x^{0}+y^{0})} \int \frac{d^{3}p}{2E_{\bm{p}}} \left[ e^{i\bm{p}\cdot(\bm{x}' - \bm{y}')} - e^{-i\bm{p}\cdot(\bm{x}' - \bm{y}')} \right] 
\nonumber 
\\
&= \frac{1}{ (2\pi)^{3} } e^{-\frac{\gamma}{2}(x^{0}+y^{0})} \int \frac{d^{3}p}{2E_{\bm{p}}} e^{i\bm{p}\cdot(\bm{x}' - \bm{y}')} - \frac{1}{ (2\pi)^{3} } e^{-\frac{\gamma}{2}(x^{0}+y^{0})} \int \frac{d^{3}\tilde{p}}{2E_{\bm{\tilde{p}}}} e^{i\bm{\tilde{p}}\cdot(\bm{x}' - \bm{y}')} 
\nonumber 
\\
&= 0, 
\label{eq:commPhi2}
\end{align}
where 
$\bm{\tilde{p}} \equiv -\bm{p}$ in the second line.

Next, we will prove the Eqs.\eqref{eq:PiPhi}, respectively. 
In the same manner performed to prove Eq.\eqref{eq:PhiPhi},  we just use the Lorentz invariance of the measure 
$d^{3}p/2E_{\bm{p}}$ and choose the Lorentz transformation 
$\Lambda$ such that 
$x'_{0} - y'_{0} = 0$ for 
$(x^0-y^0)^2-(\bm{x}-\bm{y})^2<0$. 
Therefore, the following results will be obtained
\begin{align}
\left[ \hat{\Pi}_\text{H}(x^{0},\bm{x}), \hat{\Pi}_\text{H}(y^{0},\bm{y}) \right]
& 
= e^{-\frac{\gamma}{2}(x^{0} + y^{0})} \left[ \hat{\Pi}_\text{I}(x^{0},\bm{x}), \hat{\Pi}_\text{I}(y^{0},\bm{y}) \right]
\nonumber 
\\
&
= \frac{e^{-\frac{\gamma}{2}(x^{0} + y^{0})}}{ (2\pi)^{3} }  \int \frac{d^{3}p}{ 2E_{\bm{p}} } \cdot E^{2}_{\bm{p}} \left[ e^{ i p^{\mu}(x_{\mu} - y_{\mu}) } - e^{ -i p^{\mu}(x_{\mu} - y_{\mu})} \right]
\nonumber 
\\
&
= \frac{e^{-\frac{\gamma}{2}(x^{0} + y^{0})}}{ (2\pi)^{3} }  \int d^{4}p \ \delta(p_{\mu}p^{\mu} + m^{2})\theta(p^{0}) \ (p^{0})^{2} \left[ e^{ i p^{\mu}(x_{\mu} - y_{\mu}) } - e^{ -i p^{\mu}(x_{\mu} - y_{\mu})} \right] 
\nonumber 
\\
&
= \frac{e^{-\frac{\gamma}{2}(x^{0} + y^{0})} }{ (2\pi)^{3} }\int d^{4}p' \ \delta(p'_{\mu}p'^{\mu} + m^{2})\theta(p'^{0})(p'^{0})^{2}\left[ e^{ i p'^{\mu}(x_\mu - y_\mu) } - e^{ -i p'^{\mu}(x_\mu - y_\mu)} \right]
\nonumber 
\\
&
= 
\frac{ e^{-\frac{\gamma}{2}(x^{0} + y^{0})} }{(2\pi)^{3}}
\int d^{4}p \ \delta(p_{\mu}p^{\mu} + m^{2})\theta(p^{0}) \ (\Lambda^{0}_{\ \nu}p^{\nu})^{2} 
\left[ e^{ ip^{\mu}(x'_{\mu} - y'_{\mu}) } - e^{ -ip^{\mu}(x'_{\mu} - y'_{\mu}) } \right]
\nonumber 
\\
&
= 
\frac{ e^{-\frac{\gamma}{2}(x^{0} + y^{0})} }{(2\pi)^{3}}
\int \frac{d^3p}{2E_{\bm{p}}} (\Lambda^0{}_0 E_{\bm{p}}+\Lambda^0{}_j p^{j})^{2} 
\left[
e^{i \bm{p}\cdot( \bm{x}' - \bm{y}')} 
- e^{-i \bm{p}\cdot( \bm{x}' - \bm{y}')} \right]
\nonumber 
\\
&
=  
\frac{e^{-\frac{\gamma}{2}(x^{0} + y^{0})} }{ (2\pi)^{3} } 
\int \frac{ d^{3}p }{ 2E_{\bm{p}} } \cdot 2\Lambda^{0}_{\ 0} \Lambda^{0}_{\ j} E_{\bm{p}}p^{j}
\left[
e^{i \bm{p}\cdot( \bm{x}' - \bm{y}')} 
- e^{-i \bm{p}\cdot( \bm{x}' - \bm{y}')} \right]
\nonumber 
\\
&
=\frac{e^{-\frac{\gamma}{2}(x^{0} + y^{0})}}{(2\pi)^3}  \Lambda^{0}_{\ 0} \Lambda^{0}_{\ j}  \int d^{3}p \ (-i) \frac{\partial}{\partial x'_{j}} \left[ e^{i \bm{p}\cdot( \bm{x}' - \bm{y}')} + e^{-i \bm{p}\cdot( \bm{x}' - \bm{y}')} \right]
\nonumber 
\\
&
=-2i e^{-\frac{\gamma}{2}(x^{0} + y^{0})} \ \Lambda^{0}_{\ 0} \Lambda^{0}_{\ j} \frac{\partial}{\partial x'_{j}} \delta^{3}(\bm{x}' - \bm{y}') = 0, 
\label{eq:comm.PiPi} 
\end{align}
where we used the fact that the vectors 
$\bm{x}'$
and
$\bm{y}'$ are space-like separated since the vectors 
$x^\mu$ and 
$y^\mu$ are space-like separated.
We further show that 
\begin{align}
\left[ \hat{\Phi}_\text{H}(x^{0},\bm{x}), \hat{\Pi}_\text{H}(y^{0},\bm{y}) \right]
&= e^{-\frac{\gamma}{2}(x^{0} + y^{0})} \left[ \hat{\Phi}_\text{I}(x^{0},\bm{x}), \hat{\Pi}_\text{I}(y^{0},\bm{y}) \right]
\nonumber
\\
&= \frac{i e^{-\frac{\gamma}{2}(x^{0} + y^{0})} }{(2\pi)^{3}}\int \frac{d^{3}p}{ 2E_{\bm{p}} } \, E_{\bm{p}} \left[ e^{ i p^{\mu}(x_\mu -y_\mu) } + e^{ -i p^{\mu}(x_{\mu} - y_{\mu})} \right]
\nonumber
\\
&= \frac{ie^{-\frac{\gamma}{2}(x^{0} + y^{0})}}{ (2\pi)^3} \int d^{4}p \ \delta(p_{\mu}p^{\mu} + m^{2})\theta(p^{0}) \ p^{0}
\left[ e^{ i p^{\mu}(x_{\mu} - y_{\mu}) } + e^{ -i p^{\mu}(x_{\mu} - y_{\mu})} \right] 
\nonumber
\\
&= \frac{ie^{-\frac{\gamma}{2}(x^{0} + y^{0})}}{ (2\pi)^{3} }  \int d^{4}p \ \delta(p_\mu p^\mu + m^{2})\theta(p^{0}) \ \Lambda^{0}_{\ \nu} p^{\nu} 
\left[ e^{ i p^{\rho}(x'_{\rho} - y'_{\rho}) } + e^{ -i p^{\rho}(x'_{\rho} - y'_{\rho})} \right]
\nonumber
\\
&= \frac{ie^{-\frac{\gamma}{2}(x^{0} + y^{0})} }{ (2\pi)^{3} } \int \frac{d^{3}p}{ 2E_{\bm{p}} }  \Lambda^{0}_{\ 0} E_{\bm{p}}
\left[ e^{i \bm{p}\cdot( \bm{x}' - \bm{y}')} + e^{-i \bm{p}\cdot( \bm{x}' - \bm{y}')} \right]
\nonumber
\\
&= i \Lambda^{0}_{\ 0} \ e^{-\frac{\gamma}{2}(x^{0} + y^{0})} \delta^{3}(\bm{x}' - \bm{y}') = 0, 
\label{eq:comm.PhiPi}
\end{align}
where we used the fact that the vectors 
$x^\mu$ and 
$y^\mu$ are space-like separated. 
We conclude the proofs of the three commutation relations.

\section{Time evolution of the product of operators}
\label{sec:prod}

In Sec.\ref{sec:model}, we stated that any operators commutate if they are space-like separated under the model of the master equation derived in this study. 
We prove this statement in this section. To this end, we have to find the time evolution of the product $\hat{\xi}^{i'_{1}}(\bm{x}'_{1})\cdots \hat{\xi}^{i'_{N}}(\bm{x}'_{N}) \hat{\xi}^{i_{1}}(\bm{x}_{1}) \cdots \hat{\xi}^{i_{M}}(\bm{x}_{M})$. With reference Eqs.\eqref{eq:PhiI} and \eqref{eq:HI}, the following relation is satisfied:
\begin{equation}
    e^{\mathcal{L}^{\dagger}t} \left[ \hat{\xi}^{i'_{1}}(\bm{x}'_{1})\cdots \hat{\xi}^{i'_{N}}(\bm{x}'_{N}) \hat{\xi}^{i_{1}}(\bm{x}_{1}) \cdots \hat{\xi}^{i_{M}}(\bm{x}_{M}) \right]
    = e^{\mathcal{D}^{\dagger}t} \left[ \hat{\xi}_{\mathrm{I}}^{i'_{1}}(t,\bm{x}'_{1})\cdots \hat{\xi}_{\mathrm{I}}^{i'_{N}}(t,\bm{x}'_{N}) \hat{\xi}_{\mathrm{I}}^{i_{1}}(t,\bm{x}_{1}) \cdots \hat{\xi}_{\mathrm{I}}^{i_{M}}(t,\bm{x}_{M}) \right] 
    \label{eq:HI2}
\end{equation}
If the action 
$(\mathcal{D}^{\dagger})^{k}$ can be computed, the right hand side of \eqref{eq:HI2} is obtained. 
We introduce the normal ordered product and use the Wick's theorem to calculate the right hand side of \eqref{eq:HI2}.

The operator 
$\hat{\xi}^{i}_{\mathrm{I}}(t,\bm{x})$ is decomposed into the term which contains annihilation operators 
$\hat{\zeta}^{i}_{+}(t,\bm{x})$ and the term which contains creation operators 
$\hat{\zeta}^{i}_{-}(t,\bm{x})$:
\begin{equation}
\hat{\xi}^{i}_{\mathrm{I}}(t,\bm{x}) = \hat{\zeta}^{i}_{+}(t,\bm{x}) + \hat{\zeta}^{i}_{-}(t,\bm{x}),
\label{eq:decomp.}
\end{equation}
where 
$\hat{\zeta}^{i}_{+}(t,\bm{x}_{i})$ and 
$\hat{\zeta}^{i}_{-}(t,\bm{x}_{i})$ are called the positive frequency term and the negative frequency term, respectively. 
For two field operators 
$\hat{\xi}^{i}_{\mathrm{I}}(t,\bm{x})$ and 
$\hat{\xi}^{j}_{\mathrm{I}}(t,\bm{y})$, the following product 
$\mathcal{N}[ \hat{\xi}^{i}_\text{I}(t,\bm{x})\hat{\xi}^{j}_\text{I}(t,\bm{y}) ]$ is called the normal ordered product,
\begin{equation}
\mathcal{N}\left[ \hat{\xi}^{i}_{\mathrm{I}}(t,\bm{x})\hat{\xi}^{j}_{\mathrm{I}}(t,\bm{y}) \right] = \hat{\zeta}^{i}_{+}(t,\bm{x})\hat{\zeta}^{j}_{+}(t,\bm{y}) + \hat{\zeta}^{j}_{-}(t,\bm{y})\hat{\zeta}^{i}_{+}(t,\bm{x}) + \hat{\zeta}^{i}_{-}(t,\bm{x})\hat{\zeta}^{j}_{+}(t,\bm{y}) + \hat{\zeta}^{i}_{-}(t,\bm{x})\hat{\zeta}^{j}_{-}(t,\bm{y}).
\label{eq:normal}
\end{equation}
The convenient point of the normal ordered product is that the calculation of 
$\mathcal{D}^{\dagger}$ is simple. 
The normal ordered product 
$\mathcal{N}[ \hat{\xi}^{i_{1}}_{\mathrm{I}}(t,\bm{x}_{1})\hat{\xi}^{i_{2}}_{\mathrm{I}}(t,\bm{x}_{2}) \cdots \hat{\xi}^{i_{n}}_{\mathrm{I}}(t,\bm{x}_{n}) ]$ of the 
$n$ fields is
\begin{align}
&\mathcal{N}[ \hat{\xi}^{i_{1}}_{\mathrm{I}}(t,\bm{x}_{1}) \cdots \hat{\xi}^{i_{n}}_{\mathrm{I}}(t,\bm{x}_{n}) ]
\nonumber 
\\
&\quad 
= \hat{\zeta}^{i_{1}}_{+}(t,\bm{x}_{1})\hat{\zeta}^{i_{2}}_{+}(t,\bm{x}_{2}) \cdots \hat{\zeta}^{i_{n}}_{+}(t,\bm{x}_{n})
\nonumber 
\\
&\quad \quad
+\hat{\zeta}^{i_{1}}_{-}(t,\bm{x}_{1})\hat{\zeta}^{i_{2}}_{+}(t,\bm{x}_{2})\hat{\zeta}^{i_{3}}_{+}(t,\bm{x}_{3}) \cdots \hat{\zeta}^{i_{n}}_{+}(t,\bm{x}_{n})  
+ \hat{\zeta}^{i_{2}}_{-}(t,\bm{x}_{2})\hat{\zeta}^{i_{1}}_{+}(t,\bm{x}_{1})\hat{\zeta}^{i_{3}}_{+}(t,\bm{x}_{3}) \cdots \hat{\zeta}^{i_{n}}_{+}(t,\bm{x}_{n}) 
+ \cdots 
\nonumber 
\\
&\quad \quad + \hat{\zeta}^{i_{1}}_{-}(t,\bm{x}_{1})\hat{\zeta}^{i_{2}}_{-}(t,\bm{x}_{2})\hat{\zeta}^{i_{3}}_{+}(t,\bm{x}_{3}) \cdots \hat{\zeta}^{i_{n}}_{+}(t,\bm{x}_{n}) 
+ \hat{\zeta}^{i_{1}}_{-}(t,\bm{x}_{1})\hat{\zeta}^{i_{3}}_{-}(t,\bm{x}_{3})\hat{\zeta}^{i_{2}}_{+}(t,\bm{x}_{2}) \cdots \hat{\zeta}^{i_{n}}_{+}(t,\bm{x}_{n}) 
+ \cdots 
\nonumber 
\\
& \hspace*{6.0cm} \vdots 
\nonumber 
\\
& \quad \quad  
+ \hat{\zeta}^{i_{1}}_{-}(t,\bm{x}_{1}) \cdots 
\hat{\zeta}^{i_{n-1}}_{-}(t,\bm{x}_{n-1})\hat{\zeta}^{i_{n}}_{+}(t,\bm{x}_{n}) 
+ \hat{\zeta}^{i_{1}}_{-}(t,\bm{x}_{1}) \cdots 
\hat{\zeta}^{i_{n}}_{-}(t,\bm{x}_{n})\hat{\zeta}^{i_{n-1}}_{+}(t,\bm{x}_{n-1}) 
+ \cdots 
\nonumber 
\\
& \quad \quad 
+ \hat{\zeta}^{i_{1}}_{-}(t,\bm{x}_{1})\hat{\zeta}^{i_{2}}_{-}(t,\bm{x}_{2}) \cdots \hat{\zeta}^{i_{n}}_{-}(t,\bm{x}_{n}).
\label{eq:NOn}
\end{align}
To make 
$\mathcal{D}^{\dagger}$ work for 
$\mathcal{N}[ \hat{\xi}^{i_{1}}_{\mathrm{I}}(t,\bm{x}_{1})\hat{\xi}^{i_{2}}_{\mathrm{I}}(t,\bm{x}_{2}) \cdots \hat{\xi}^{i_{n}}_{\mathrm{I}}(t,\bm{x}_{n}) ]$, we can use the following relation:If the product with 
$k$ pieces of 
$\hat{\zeta}_{-}$ lined up on the left, the action of $\mathcal{D}^{\dagger}$ on that is
\begin{equation}
    \mathcal{D}^{\dagger}[ \underbrace{\hat{\zeta}_{-} \cdots \hat{\zeta}_{-}}_{k \ \mathrm{pieces}} \underbrace{\hat{\zeta}_{+} \cdots \hat{\zeta}_{+}}_{n-k \ \mathrm{pieces}} ] = - \frac{n}{2}\gamma \underbrace{\hat{\zeta}_{-} \cdots \hat{\zeta}_{-}}_{k \ \mathrm{pieces}} \underbrace{\hat{\zeta}_{+} \cdots \hat{\zeta}_{+}}_{n-k \ \mathrm{pieces}}. 
    \label{eq:DaggerNO}
\end{equation}
Eq.\eqref{eq:DaggerNO} holds for any $k$, so $\mathcal{D}^{\dagger}[ \mathcal{N}[ \hat{\xi}^{i_{1}}_{\mathrm{I}}(t,\bm{x}_{1})\hat{\xi}^{i_{2}}_{\mathrm{I}}(t,\bm{x}_{2}) \cdots \hat{\xi}^{i_{n}}_{\mathrm{I}}(t,\bm{x}_{n}) ] ]$ is computed as
\begin{equation}
    \mathcal{D}^{\dagger}\left[ \mathcal{N}\left[ \hat{\xi}^{i_{1}}_{\mathrm{I}}(t,\bm{x}_{1})\hat{\xi}^{i_{2}}_{\mathrm{I}}(t,\bm{x}_{2}) \cdots \hat{\xi}^{i_{n}}_{\mathrm{I}}(t,\bm{x}_{n}) \right] \right] = - \frac{n}{2}\gamma \ \mathcal{N}\left[ \hat{\xi}^{i_{1}}_{\mathrm{I}}(t,\bm{x}_{1})\hat{\xi}^{i_{2}}_{\mathrm{I}}(t,\bm{x}_{2}) \cdots \hat{\xi}^{i_{n}}_{\mathrm{I}}(t,\bm{x}_{n}) \right].
    \label{eq:DaggerNO2}
\end{equation}
Since what we originally want to know is 
$\mathcal{D}^{\dagger}[\hat{\xi}^{i_{1}}_{\mathrm{I}}(t,\bm{x}_{1})\hat{\xi}^{i_{2}}_{\mathrm{I}}(t,\bm{x}_{2}) \cdots \hat{\xi}^{i_{n}}_{\mathrm{I}}(t,\bm{x}_{n}) ]$, the relation between $\mathcal{D}^{\dagger}[ \mathcal{N}[ \hat{\xi}^{i_{1}}_{\mathrm{I}}(t,\bm{x}_{1})\hat{\xi}^{i_{2}}_{\mathrm{I}}(t,\bm{x}_{2}) \cdots \hat{\xi}^{i_{n}}_{\mathrm{I}}(t,\bm{x}_{n}) ] ]$ and $\mathcal{D}^{\dagger}[  \hat{\xi}^{i_{1}}_{\mathrm{I}}(t,\bm{x}_{1})\hat{\xi}^{i_{2}}_{\mathrm{I}}(t,\bm{x}_{2}) \cdots \hat{\xi}^{i_{n}}_{\mathrm{I}}(t,\bm{x}_{n}) ]$ is needed. 
The relation is given by Wick's theorem as follows: the product $\hat{\xi}^{i_{1}}_{\mathrm{I}}(t,\bm{x}_{1})\hat{\xi}^{i_{2}}_{\mathrm{I}}(t,\bm{x}_{2}) \cdots \hat{\xi}^{i_{n}}_{\mathrm{I}}(t,\bm{x}_{n})$ is represented by normal ordered product, 
\begin{align}
    \hat{\xi}^{i_{1}}_{\mathrm{I}}(t,\bm{x}_{1})\hat{\xi}^{i_{2}}_{\mathrm{I}}(t,\bm{x}_{2}) \cdots \hat{\xi}^{i_{n}}_{\mathrm{I}}(t,\bm{x}_{n}) 
    &= \mathcal{N}[ \hat{\xi}^{i_{1}}_{\mathrm{I}}(t,\bm{x}_{1})\hat{\xi}^{i_{2}}_{\mathrm{I}}(t,\bm{x}_{2}) \cdots \hat{\xi}^{i_{n}}_{\mathrm{I}}(t,\bm{x}_{n}) ] 
    \nonumber \\
    &\quad + \sum_{1-\mathrm{pair}} \mathcal{N}[ \hat{\xi}^{i_{1}}_{\mathrm{I}}(t,\bm{x}_{1})\hat{\xi}^{i_{2}}_{\mathrm{I}}(t,\bm{x}_{2}) \contraction[1ex]{}{\cdots}{}{\cdots} \cdots\cdots\cdots\cdots \hat{\xi}^{i_{n}}_{\mathrm{I}}(t,\bm{x}_{n}) ]
    \nonumber \\
    & \quad + \sum_{2-\mathrm{paris}}  \mathcal{N}[ \hat{\xi}^{i_{1}}_{\mathrm{I}}(t,\bm{x}_{1})\hat{\xi}^{i_{2}}_{\mathrm{I}}(t,\bm{x}_{2}) \contraction{}{\cdots}{\cdots}{\cdots}\contraction[2ex]{\cdots}{\cdots}{\cdots}{\cdots}\cdots\cdots\cdots\cdots \cdots\ \hat{\xi}^{i_{n}}_{\mathrm{I}}(t,\bm{x}_{n}) ] 
    \nonumber \\
    & \quad + (3-\mathrm{pairs \ or \ more \ terms}) \cdots,
    \label{eq:Wick}
\end{align}
where 
$\contraction{}{\cdots}{}{\cdots}\cdots\cdots$ is called the Wick contraction and is given as the expectation value taken in a vacuum for the product of two operators connected by the line. 
Considering, for example, the case of the product $\hat{\xi}^{i}_{\mathrm{I}}(t,\bm{x})\hat{\xi}^{j}_{\mathrm{I}}(t,\bm{y})$, the Wick's theorem gives
\begin{align}
    \hat{\xi}^{i}_{\mathrm{I}}(t,\bm{x})\hat{\xi}^{j}_{\mathrm{I}}(t,\bm{y}) 
    &= \mathcal{N} \left[ \hat{\xi}^{i}_{\mathrm{I}}(t,\bm{x})\hat{\xi}^{j}_{\mathrm{I}}(t,\bm{y}) \right] + \contraction{}{ \hat{\xi}^{i}_{\mathrm{I}}(t,\bm{x}) }{}{ \hat{\xi}^{j}_{\mathrm{I}}(t,\bm{y}) }\hat{\xi}^{i}_{\mathrm{I}}(t,\bm{x})\hat{\xi}^{j}_{\mathrm{I}}(t,\bm{y}) 
    \nonumber 
    \\
    &= \mathcal{N} \left[ \hat{\xi}^{i}_{\mathrm{I}}(t,\bm{x})\hat{\xi}^{j}_{\mathrm{I}}(t,\bm{y}) \right] + \langle 0| \hat{\xi}^{i}_{\mathrm{I}}(t,\bm{x})\hat{\xi}^{j}_{\mathrm{I}}(t,\bm{y}) | 0 \rangle.
    \label{eq:Norm.ex.}
\end{align}
Since 
$\mathcal{D}^{\dagger}$ is a linear superoperator, using \eqref{eq:DaggerNO2} and \eqref{eq:Wick}, 
$\mathcal{D}^{\dagger}[\hat{\xi}^{i_{1}}_{\mathrm{I}}(t,\bm{x}_{1})\hat{\xi}^{i_{2}}_{\mathrm{I}}(t,\bm{x}_{2}) \cdots \hat{\xi}^{i_{n}}_{\mathrm{I}}(t,\bm{x}_{n})] $ 
is computed as
\begin{align}
\mathcal{D}^{\dagger}
\left[
\hat{\xi}^{i_{1}}_{\mathrm{I}}(t,\bm{x}_{1})\hat{\xi}^{i_{2}}_{\mathrm{I}}(t,\bm{x}_{2}) 
\cdots \hat{\xi}^{i_{n}}_{\mathrm{I}}(t,\bm{x}_{n}) 
\right]
&=\mathcal{D}^{\dagger}\left[ \mathcal{N}[ \hat{\xi}^{i_{1}}_{\mathrm{I}}(t,\bm{x}_{1})\hat{\xi}^{i_{2}}_{\mathrm{I}}(t,\bm{x}_{2}) \cdots \hat{\xi}^{i_{n}}_{\mathrm{I}}(t,\bm{x}_{n}) ] \right] 
\nonumber 
\\
&+ \sum_{1-\mathrm{pair}} \mathcal{D}^{\dagger} \left[ \mathcal{N}[ \hat{\xi}^{i_{1}}_{\mathrm{I}}(t,\bm{x}_{1})\hat{\xi}^{i_{2}}_{\mathrm{I}}(t,\bm{x}_{2}) \contraction[1ex]{}{\cdots}{}{\cdots} \cdots\cdots\cdots\cdots \hat{\xi}^{i_{n}}_{\mathrm{I}}(t,\bm{x}_{n}) ] \right] 
\nonumber 
\\
&+ \sum_{2-\mathrm{pairs}} \mathcal{D}^{\dagger} 
\left[ 
\mathcal{N}[ \hat{\xi}^{i_{1}}_{\mathrm{I}}(t,\bm{x}_{1})\hat{\xi}^{i_{2}}_{\mathrm{I}}(t,\bm{x}_{2}) \contraction{}{\cdots}{\cdots}{\cdots}\contraction[2ex]{\cdots}{\cdots}{\cdots}{\cdots}\cdots\cdots\cdots\cdots \cdots\ \hat{\xi}^{i_{n}}_{\mathrm{I}}(t,\bm{x}_{n}) 
] 
\right] 
\nonumber 
\\
&
+ ( 3-\mathrm{ pairs \ or \ more \ terms} )
\nonumber 
\\
&= - \frac{n}{2} \gamma \ \mathcal{N} \left[ \hat{\xi}^{i_{1}}_{\mathrm{I}}(t,\bm{x}_{1})\hat{\xi}^{i_{2}}_{\mathrm{I}}(t,\bm{x}_{2}) \cdots \hat{\xi}^{i_{n}}_{\mathrm{I}}(t,\bm{x}_{n}) \right] 
\nonumber 
\\
&- \sum_{1-\mathrm{pair}}\frac{n-2}{2} \gamma \ \mathcal{N} \left[ \hat{\xi}^{i_{1}}_{\mathrm{I}}(t,\bm{x}_{1})\hat{\xi}^{i_{2}}_{\mathrm{I}}(t,\bm{x}_{2}) \contraction[1ex]{}{\cdots}{}{\cdots} \cdots\cdots\cdots\cdots \hat{\xi}^{i_{n}}_{\mathrm{I}}(t,\bm{x}_{n}) \right] 
\nonumber 
\\
& 
 -\sum_{2-\mathrm{pairs}} \frac{n-4}{2} \gamma \ 
\mathcal{N}\left[
\hat{\xi}^{i_{1}}_{\mathrm{I}}(t,\bm{x}_{1})\hat{\xi}^{i_{2}}_{\mathrm{I}}(t,\bm{x}_{2}) \contraction{}{\cdots}{\cdots}{\cdots}\contraction[2ex]{\cdots}{\cdots}{\cdots}{\cdots}\cdots\cdots\cdots\cdots \cdots\ \hat{\xi}^{i_{n}}_{\mathrm{I}}(t,\bm{x}_{n}) 
\right] 
\nonumber 
\\
&
+ ( 3-\mathrm{ pairs \ or \ more \ terms} ). 
\label{eq:Ddgxis}
\end{align}
Furthermore, $(\mathcal{D}^{\dagger})^{k}$ operates as,
\begin{align}
&(\mathcal{D}^{\dagger})^{k}\left[ \hat{\xi}^{i_{1}}_{\mathrm{I}}(t,\bm{x}_{1})\hat{\xi}^{i_{2}}_{\mathrm{I}}(t,\bm{x}_{2}) \cdots \hat{\xi}^{i_{n}}_{\mathrm{I}}(t,\bm{x}_{n}) \right] 
\nonumber 
\\
&\qquad \qquad= \left(- \frac{n}{2} \gamma\right)^{k} \ \mathcal{N} \left[ \hat{\xi}^{i_{1}}_{\mathrm{I}}(t,\bm{x}_{1})\hat{\xi}^{i_{2}}_{\mathrm{I}}(t,\bm{x}_{2}) \cdots \hat{\xi}^{i_{n}}_{\mathrm{I}}(t,\bm{x}_{n}) \right] 
\nonumber 
\\
& \qquad \qquad + \sum_{1-\mathrm{pair}} \left( - \frac{n-2}{2} \gamma\right)^{k} \ \mathcal{N} \left[ \hat{\xi}^{i_{1}}_{\mathrm{I}}(t,\bm{x}_{1})\hat{\xi}^{i_{2}}_{\mathrm{I}}(t,\bm{x}_{2}) \contraction[1ex]{}{\cdots}{}{\cdots} \cdots\cdots\cdots\cdots\hat{\xi}^{i_{n}}_{\mathrm{I}}(t,\bm{x}_{n}) \right] 
\nonumber 
\\
&\qquad \qquad + \sum_{2-\mathrm{pairs}} \left(-\frac{n-4}{2} \gamma\right)^{k} \ \mathcal{N}\left[ \hat{\xi}^{i_{1}}_{\mathrm{I}}(t,\bm{x}_{1})\hat{\xi}^{i_{2}}_{\mathrm{I}}(t,\bm{x}_{2}) \contraction{}{\cdots}{\cdots}{\cdots}\contraction[2ex]{\cdots}{\cdots}{\cdots}{\cdots}\cdots\cdots\cdots\cdots \cdots\ \hat{\xi}^{i_{n}}_{\mathrm{I}}(t,\bm{x}_{n}) \right]
\nonumber 
\\
& \qquad \qquad + ( 3-\mathrm{ pairs \ or \ more \ terms} ). 
\label{eq:Ddgxis2}
\end{align}
Therefore, we can calculate the Heisenberg operator $ e^{ \mathcal{L}^{\dagger}t }[ \hat{\xi}^{i_{1}}(\bm{x}_{1})\hat{\xi}^{i_{2}}(\bm{x}_{2}) \cdots \hat{\xi}^{i_{n}}(\bm{x}_{n}) ] $ because of Eqs.\eqref{eq:HI2} and \eqref{eq:Ddgxis2}:
\begin{align}
    &e^{ \mathcal{L}^{\dagger}t }\left[ \hat{\xi}^{i_{1}}(\bm{x}_{1})\hat{\xi}^{i_{2}}(\bm{x}_{2}) \cdots \hat{\xi}^{i_{n}}(\bm{x}_{n}) \right]
    \nonumber 
    \\
    &\qquad \qquad = \exp( - \frac{n}{2}\gamma t ) \ \mathcal{N} \left[ \hat{\xi}^{i_{1}}_{\mathrm{I}}(t,\bm{x}_{1})\hat{\xi}^{i_{2}}_{\mathrm{I}}(t,\bm{x}_{2}) \cdots \hat{\xi}^{i_{n}}_{\mathrm{I}}(t,\bm{x}_{n}) \right] \notag 
    \\
    & \qquad \qquad + \exp( -\frac{n-2}{2}\gamma t ) \ \sum_{1-\mathrm{pair}}\mathcal{N} \left[ \hat{\xi}^{i_{1}}_{\mathrm{I}}(t,\bm{x}_{1})\hat{\xi}^{i_{2}}_{\mathrm{I}}(t,\bm{x}_{2}) \contraction[1ex]{}{\cdots}{}{\cdots} \cdots\cdots\cdots\cdots \hat{\xi}^{i_{n}}_{\mathrm{I}}(t,\bm{x}_{n}) \right] \notag 
    \\
    & \qquad \qquad + \exp( - \frac{n-4}{2}\gamma t ) \ \sum_{2-\mathrm{pair}}\mathcal{N}\left[ \hat{\xi}^{i_{1}}_{\mathrm{I}}(t,\bm{x}_{1})\hat{\xi}^{i_{2}}_{\mathrm{I}}(t,\bm{x}_{2}) \contraction{}{\cdots}{\cdots}{\cdots}\contraction[2ex]{\cdots}{\cdots}{\cdots}{\cdots}\cdots\cdots\cdots\cdots \cdots\ \hat{\xi}^{i_{n}}_{\mathrm{I}}(t,\bm{x}_{n}) \right] \notag 
    \\
    & \qquad \qquad + ( 3-\mathrm{ pairs \ or \ more \ terms} ) \notag 
    \\
    &\qquad \qquad =  \mathcal{N} \left[ \hat{\xi}^{i_{1}}_{\mathrm{H}}(t,\bm{x}_{1})\hat{\xi}^{i_{2}}_{\mathrm{H}}(t,\bm{x}_{2}) \cdots \hat{\xi}^{i_{n}}_{\mathrm{H}}(t,\bm{x}_{n}) \right] \notag 
    \\
    & \qquad \qquad +  \exp(\gamma t) \sum_{1-\mathrm{pair}}\mathcal{N} \left[ \hat{\xi}^{i_{1}}_{\mathrm{H}}(t,\bm{x}_{1})\hat{\xi}^{i_{2}}_{\mathrm{H}}(t,\bm{x}_{2}) \contraction[1ex]{}{\cdots}{}{\cdots} \cdots\cdots\cdots\cdots \hat{\xi}^{i_{n}}_{\mathrm{H}}(t,\bm{x}_{n}) \right] \notag \\
    & \qquad \qquad + \exp(2\gamma t) \sum_{2-\mathrm{pair}}\mathcal{N}\left[ \hat{\xi}^{i_{1}}_{\mathrm{H}}(t,\bm{x}_{1})\hat{\xi}^{i_{2}}_{\mathrm{H}}(t,\bm{x}_{2}) \contraction{}{\cdots}{\cdots}{\cdots}\contraction[2ex]{\cdots}{\cdots}{\cdots}{\cdots}\cdots\cdots\cdots\cdots\cdots \hat{\xi}^{i_{n}}_{\mathrm{H}}(t,\bm{x}_{n}) \right] \notag \\
    & \qquad \qquad + ( 3-\mathrm{ pairs \ or \ more \ terms} ),
    \label{eq:HI3}
\end{align}
where we used the relation Eq.\eqref{eq:PhiH2} in the last equality. 
The factors, such as $\exp(\gamma t)$ and $\exp(2\gamma t)$, appear since the Wick contraction of  
$\hat{\xi}^{i}_{\mathrm{I}}$ connects with 
that of 
$\hat{\xi}^{i}_{\mathrm{H}}$ as
\begin{equation}
    \contraction{}{ \hat{\xi}^{i}_{\mathrm{I}}(t,\bm{x}) }{}{ \hat{\xi}^{j}_{\mathrm{I}}(t,\bm{y}) }\hat{\xi}^{i}_{\mathrm{I}}(t,\bm{x})\hat{\xi}^{j}_{\mathrm{I}}(t,\bm{y}) =\exp(\gamma t)\contraction{}{ \hat{\xi}^{i}_{\mathrm{H}}(t,\bm{x}) }{}{ \hat{\xi}^{j}_{\mathrm{I}}(t,\bm{y}) }\hat{\xi}^{i}_{\mathrm{H}}(t,\bm{x})\hat{\xi}^{j}_{\mathrm{H}}(t,\bm{y}).
    \label{eq:wicks}
\end{equation}
The result \eqref{eq:HI3} means that the time evolution $e^{ \mathcal{L}^{\dagger}t }[ \hat{\xi}^{i_{1}}(\bm{x}_{1})\hat{\xi}^{i_{2}}(\bm{x}_{2}) \cdots \hat{\xi}^{i_{n}}(\bm{x}_{n}) ]$ is written with the combinations of the products of the Heisenberg operators chosen from $\hat{\xi}^{i_{1}}_{\mathrm{H}}(t,\bm{x}_{1}),...., \hat{\xi}^{i_{n}}_{\mathrm{H}}(t,\bm{x}_{n})$ by repeatedly using the Wick's theorem. 

The above result implies that the commutation relation of the arbitrary two operators can be calculated from the commutation relations $[ \hat{\xi}^{i_{k}}_{\mathrm{H}}, \hat{\xi}^{j_{l}}_{\mathrm{H}} ]$. Therefore, supposing that $\hat{A}$ is a operator defined for a spatial region A at time $t = 0$ and $\hat{B}$ is a operator defined for a spatial region B at time $t' = 0$, it turns out that the time-evolved operators $\hat{A}_{\mathrm{H}}(t)$ and $\hat{B}_{\mathrm{H}}(t')$ commutes if they are spacelike separated.

\section{Derivation of the Markovian quantum master equation Eq.\eqref{eq:massive}}
\label{app:massive}

We assume that the massive spectrum of 
$\hat{P}^{\mu}$ satisfies
\begin{equation}
\hat{P}^{\mu}\hat{P}_{\mu} = - m^{2}, \ \hat{P}^{0} > 0. \label{eq:spec}
\end{equation}
The equation \eqref{eq:spec} leads to the Hamiltonian 
$\hat{H} = \hat{P}^{0}$ given as 
$\hat{H} = \sqrt{\hat{P}^{k}\hat{P}_{k} + m^2}$. 
In this appendix, thinking about the cases in Table \ref{tab:1}, we derive the Lindblad operator 
$\hat{L}_{q,\xi}$ and the self-adjoint operator 
$\hat{M}$, which give the Markovian QME Eq.\eqref{eq:model}.

\textbf{\underline{Lindblad operator for the case 
 $\, \ell^\mu=[\pm M,0,0,0], \, M>0$:}}
Substituting 
$a^{\mu} = (0,\bm{a})$ into Eq.\eqref{eq:Tf}, this equation leads to 
\begin{equation}
f_{\ell,\xi}(\bm{p})e^{-i \bm{p} \cdot \bm{a}} = f_{\ell,\xi}(\bm{p}) 
\, \therefore \, 
f_{\ell,\xi}(\bm{p}) = f_{\ell,\xi}\delta^{3}(\bm{p}). 
\label{eq:fI}
\end{equation}
Eq.\eqref{eq:fI} can be checked by integrating both sides. 
Next, we think \eqref{eq:Tf} in the case of 
$a^{\mu} = (a,\bm{0})$, which leads to  
\begin{equation}
f_{\ell,\xi}(\bm{p}) e^{iE_{\bm{p}}a} = f_{\ell,\xi}(\bm{p})e^{ \pm i M a}.
\label{eq:fI2}
\end{equation}
This equation has a meaning only when 
$\bm{p} = \bm{0}$ because of \eqref{eq:fI}.
So, we eventually get
\begin{equation}
f_{\ell,\xi} e^{ima} = f_{\ell,\xi}  e^{\pm iMa}. 
\label{eq:fI3}
\end{equation}
Since the mass 
$m$ is positive, to get a nontrivial solution as
$f_{\ell,\xi}\neq 0$, we should choose $+M$ with 
$M = m$.
The above analysis implies that the Lindblad operator 
$\hat{L}_{\ell,\xi}$ with 
$\ell^\mu=[m,0,0,0]$ has 
\begin{equation}
\hat{L}_{\ell,\xi} = \int d^{3}p \ f_{\ell,\xi} \hat{a}(\bm{p}) \delta^{3}(\bm{p}) = f_{\ell,\xi}\hat{a}(\bm{0}) 
\label{eq:LlxiI}
\end{equation}
Eq.\eqref{eq:VSL} tells us that 
\begin{equation}
\hat{L}_{q,\xi} = N^*_{q} \hat{V}(S_{q}) \hat{L}_{\ell,\xi}\hat{V}^{\dagger}(S_{q}) = N^{\ast}_{q} f_{\ell,\xi} \sqrt{ \frac{ E_{\bm{q}} }{m} } \hat{a}(\bm{q}) = f_{\ell,\xi}\hat{a}(\bm{q}),
\label{eq:LqxiI}
\end{equation}
where 
$E_{\bm{q}} = (S_q \ell)^{0}$, $q^{i} = (S_q \ell)^{i}$ and 
$N_{q} = \sqrt{ m / E_{\bm{q}} }$. 
The normalization $N_{q}$ is given by setting the inner product $\bm{v}^{\dagger}_{q',\xi'}\bm{v}_{q,\xi}$ as
\begin{equation}
\bm{v}^{\dagger}_{q',\xi'}\bm{v}_{q,\xi} = \delta^{3}(\bm{q}' - \bm{q})\delta_{\xi'\xi}.
\label{eq:vinn}
\end{equation}
Under this inner product, we can get the following completeness condition,
\begin{equation}
\int d^{3}q \sum_{\xi} \bm{v}_{q,\xi}\bm{v}^{\dagger}_{q,\xi} = I.
\label{eq:complete}
\end{equation}
We then derive a part of 
$\mathcal{D}[\rho]$ as
\begin{equation}
\mathcal{D}[\rho] \supset \sum_\xi |f_{\ell,\xi}|^{2}  \int d^{3}p \left[ \hat{a}(\bm{p})\rho(t)\hat{a}^{\dagger}(\bm{p}) - \frac{1}{2}\left\{\hat{a}^{\dagger}(\bm{p})\hat{a}(\bm{p}), \rho(t)\right\}\right]. 
\label{eq:DI}
\end{equation}

\textbf{\underline{Lindblad operator for the case  
 $\, \ell^\mu=[\pm \kappa,0,0,\kappa], \, \kappa>0$:}}
Substituting $a^{\mu} = [0,\bm{a}]$ into Eq.\eqref{eq:Tf}, we have
\begin{equation}
f_{\ell,\xi}(\bm{p})e^{-i\bm{p}\cdot\bm{a}} = f_{\ell,\xi}(\bm{p})e^{-i \bm{\ell}\cdot\bm{a}} \ \therefore f_{\ell,\xi}(\bm{p}) = f_{\ell,\xi}\delta^{3}(\bm{p} - \bm{\ell}), 
\label{eq:fII}
\end{equation}
where 
$\bm{\ell} = [0,0,\kappa]^{\mathrm{T}}$. Also, substituting 
$a^{\mu} = [a,0,0,0]$ into Eq.\eqref{eq:Tf}, we obtain the following result:
\begin{equation}
f_{\ell,\xi}(\bm{p})e^{i E_{\bm{p}}a } = f_{\ell,\xi}(\bm{p})e^{\pm i\kappa a}.
\label{eq:fII2}
\end{equation}
Because of 
$f_{\ell,\xi}(\bm{p}) = f_{\ell,\xi}\delta^{3}(\bm{p} - \bm{\ell})$ , we get 
\begin{equation}
    f_{\ell,\xi}e^{i\sqrt{\kappa^{2} + m^{2}}a} = f_{\ell,\xi} e^{\pm i \kappa a} \ \therefore f_{\ell,\xi} = 0,
    \label{eq:fII3}
\end{equation}
where 
$E_{\bm{\ell}} = \sqrt{ \bm{\ell}^{2} + m^{2} } = \sqrt{ \kappa^{2} + m^{2} }$, and hence 
$f_{\ell,\xi}(\bm{p}) = 0$. 
Combined with the above analysis, the Lindblad operator 
$\hat{L}_{\ell,\xi}$ vanishies, and then
\begin{equation}
\hat{L}_{q,\xi} = N^{\ast}_{q}\hat{V}(S_{q})\hat{L}_{\ell,\xi}\hat{V}^{\dagger}(S_{q}) = 0. 
\label{eq:LqxiII}
\end{equation}

\textbf{\underline{Lindblad operator for the case 
 $\ell^\mu = [0,0,0,N]$, $N^{2}>0$:}}
Eq.\eqref{eq:Tf} for all 
$a^{\mu} = [a,0,0,0]$ leads to
\begin{equation}
f_{\ell,\xi}(\bm{p})e^{iE_{\bm{p}}a } = f_{\ell,\xi}(\bm{p}) \ \therefore f_{\ell,\xi}(\bm{p}) = 0, 
\label{eq:fIII}    
\end{equation}
where note that 
$E_{\bm{p}} = \sqrt{\bm{p}^{2} + m^{2}} \neq 0$. For this case, the Lindblad operator $\hat{L}_{\ell,\xi}$ vanishes, and we have
\begin{equation}
\hat{L}_{q,\xi} = N^{\ast}_{q}\hat{V}(S_{q})\hat{L}_{\ell,\xi}\hat{V}^{\dagger}(S_{q}) = 0. 
\label{eq:LqxiIII}
\end{equation}

\textbf{\underline{Lindblad operator for the case  
$\ell^\mu = [0,0,0,0]$:}}
For this case, Eq.\eqref{eq:Tf} for all 
$a^{\mu} = [a,0,0,0]$ is written as
\begin{equation}
f_{\xi}(\bm{p}) e^{ i E_{\bm{p}}a } = f_{\xi}(\bm{p}) \ \therefore f_{\xi}(\bm{p}) = 0,
\label{eq:fIV}
\end{equation}
where we dropped the label 
$\ell$ and note that 
$E_{\bm{p}} = \sqrt{ \bm{p}^{2} + m^{2} } \neq 0$. 
Hence, the Lindblad operator vanishes:
\begin{equation}
    \hat{L}_{\xi} = 0. \label{eq:LxiIV}
\end{equation}

\textbf{\underline{Self-adjoint operator  
$\hat{M}$:}}
Now that we can get the form of $\mathcal{D}[\rho]$, let us move to the analysis the self-adjoint operator $\hat{M} - \hat{H}$. 
Adopting 
$\Lambda = I$ and 
$a^{\mu}=[ 0, \bm{a} ]$ in Eq.\eqref{eq:g}, we can obtain
\begin{equation}
g(\bm{p}, \bm{p}') e^{i(\bm{p} - \bm{p}')\cdot \bm{a}} = g(\bm{p},\bm{p}') 
\ \therefore \ 
g(\bm{p},\bm{p}') = g(\bm{p})\delta^{3}(\bm{p} - \bm{p}'). 
\label{eq:gapp}
\end{equation}
This result can be checked by integrating the both sides. 
Also, adopting 
$a^{\mu} = 0$ and substituting \label{eq:gapp} into Eq.\eqref{eq:g}, we have
\begin{equation}
g(\bm{p}_{\Lambda}) = g(\bm{p}), 
\label{eq:gapp2}    
\end{equation}
where we used 
$E_{\bm{p}} \delta^{3}(\bm{p} - \bm{p}') = E_{\bm{p}_{\Lambda}} \delta^{3}(\bm{p}_{\Lambda} - \bm{p}'_{\Lambda})$. For 
$\bm{p} = \bm{0}$ and 
$\Lambda = S_{q}$ with 
$ (S_{q})^{\mu}_{\ \nu}k^{\nu} = q^{\mu}$ for 
$k^{\mu} = [m,0,0,0]$ in Eq.\eqref{eq:gapp2}, we obtain
\begin{equation}
g(\bm{q}) = g(\bm{0}).
\label{eq:gapp3}
\end{equation}
By defining 
$g(\bm{0})$ as 
$g$, the self-adjoint operator 
$\hat{M}$ is given as
\begin{equation}
\hat{M} = \hat{H} + g\hat{N}, \label{eq:Mapp}
\end{equation}
where 
$\hat{N}$ is the number operator \eqref{eq:N}. 

\textbf{\underline{Summary:}} Combining the above results \eqref{eq:DI}, \eqref{eq:LqxiII}, \eqref{eq:LqxiIII}, \eqref{eq:LxiIV} and \eqref{eq:Mapp}, we get the following form of the Markovian QME,
\begin{equation}
    \frac{d}{dt}\rho(t) = -i\left[\hat{H} + g\hat{N}, \rho(t)\right] + \gamma \int d^{3}p \left[ \hat{a}(\bm{p})\rho(t)\hat{a}^{\dagger}(\bm{p}) - \frac{1}{2}\left\{\hat{a}^{\dagger}(\bm{p})\hat{a}(\bm{p}), \rho(t)\right\}\right], \label{eq:modelapp}
\end{equation}
where we defined 
$\gamma$ as 
$\gamma = \sum_\xi |f_{\ell,\xi}|^{2}$.

\end{appendix}

\end{document}